\documentclass[iop, numberedappendix]{emulateapj}
\usepackage{gensymb} % degree
\usepackage{graphicx}
\usepackage{amsmath}
\usepackage{amssymb}
\usepackage{latexsym}
\usepackage{rotating}
\usepackage{url}
\usepackage{multirow}
\usepackage{enumerate} % abc enumerates
\usepackage{threeparttable}
\usepackage{longtable}
\usepackage{array} % to center cells vertically
\usepackage{float} % using [H] option for figures

\usepackage{natbib}
\usepackage[hidelinks]{hyperref}

\usepackage[usenames, dvipsnames]{color} %textcolor

\newcommand {\beq} {\begin{equation}}
\newcommand {\eeq} {\end{equation}}
\newcommand {\bdm} {\begin{displaymath}}
\newcommand {\edm} {\end{displaymath}}

\begin{document}

  \title {Circumnuclear structures in megamaser host galaxies}
  \author {Patryk Pjanka\footnotemark[1,$\dagger$], 
  Jenny E. Greene\footnotemark[1], 
  Anil C. Seth\footnotemark[2], 
  James A. Braatz\footnotemark[3], 
  Christian Henkel\footnotemark[4],
  Fred K. Y. Lo\footnotemark[3], 
  Ronald L\"{a}sker\footnotemark[5]}
  \affiliation {\footnotemark[1]Department of Astrophysical Sciences, Princeton University, 4 Ivy Lane, Princeton, NJ 08544, USA\\
  \footnotemark[2]Department of Physics and Astronomy, University of Utah, Salt Lake City, UT 84112, USA\\
  \footnotemark[3]National Radio Astronomy Observatory, 520 Edgemont Road, Charlottesville, VA 22903, USA\\
  \footnotemark[4]Max-Planck-Institut f{\"u}r Radioastronomie, Auf dem H{\"u}gel 69, D-53121 Bonn, Germany;\\ Astronomy Department, King Abdulaziz University, P.O. Box 80203, Jeddah 21589, Saudi Arabia\\
  \footnotemark[5]Finnish Centre for Astronomy with ESO (FINCA), University of Turku, V\"ais\"al\"antie 20, 21500 Kaarina, Finland}
  \date{\today}
  
 \footnotetext[$\dagger$]{E-mail: \url{ppjanka@princeton.edu}.}

 \begin{abstract}
  Using \emph{HST}, we identify circumnuclear ($100$--$500$~pc scale) structures in nine new H$_2$O megamaser host galaxies to understand the flow of matter from kpc-scale galactic structures down to the supermassive black holes (SMBHs) at galactic centers. We double the sample analyzed in a similar way by \cite{2013Greene} and consider the properties of the combined sample of 18 sources. We find that disk-like structure is virtually ubiquitous when we can resolve $<\mathrm{200}$~pc scales, in support of the notion that non-axisymmetries on these scales are a necessary condition for SMBH fueling. We perform an analysis of the orientation of our identified nuclear regions and compare it with the orientation of megamaser disks and the kpc-scale disks of the hosts. We find marginal evidence that the disk-like nuclear structures show increasing misalignment from the kpc-scale host galaxy disk as the scale of the structure decreases. In turn, we find that the orientation of both the $\sim100$~pc scale nuclear structures and their host galaxy large-scale disks is consistent with random with respect to the orientation of their respective megamaser disks.
  %In agreement with previous studies, we conclude that structures on both of these scales do not directly regulate the activity state of galactic nuclei, although they may be a necessary condition. We find nuclear spirals in 8 of our 18 sources, with five tightly-wound and three grand-design spirals. We have tentative evidence for a grand-design nuclear spiral in one galaxy without a large-scale bar (UGC~3789)}.
 \end{abstract}

 
 \maketitle
 
 \section{Introduction}\label{sect:intro}
 
 %\textcolor{red}{\bf SMBH formation problem}
 
 It is now commonly accepted that supermassive black holes with masses of $10^5-10^{10} M_{\odot}$ (SMBHs) ubiquitously reside in galactic centers (\citealt{1984Rees}, \citealt{2013Kormendy}). The evidence of their existence extends to redshifts of~$\sim7$ \citep{2011Mortlock, 2013Venemans, 2014deRosa}. How such massive black holes can form only $\sim 0.8$~Gyr after the Big Bang is still a mystery.
 
 %\textcolor{red}{\bf Possible solution: fuelling with cold gas from kpc-scales. But how does it lose its ang. mom.?}
 
 One of the main mechanisms proposed to fuel SMBHs is inflow of cold gas from the large-scale galaxy \citep{1978Heckman}. The final accretion itself is obviously facilitated on extremely small scales, where the infalling material radiates as an active galactic nucleus (AGN). However, for the gas to travel from the kpc-scale galaxy to the outer accretion structures at $\lesssim 1$~pc, more than four orders of magnitude in angular momentum must be lost. The details of this angular momentum extraction process are not yet fully understood.
 
 %\textcolor{red}{\bf $\vec{J}$ extraction at kpc-scales}
 
 On galaxy-wide scales, there are several potential mechanisms responsible for dissipation of angular momentum. Secular interactions between a (collisionless) stellar component and (collisional) gas in non-axisymmetric kpc-scale galactic structures (such as bars or spiral arms) can drive gas inwards \citep[e.g.][]{2010Hopkins}, and these interactions may be triggered in a variety of ways. During gas-rich mergers, tidal forces destabilize galactic disks to form non-axisymmetries, as seen by the numerical experiments of \citet{1989Hernquist}, \citet{1996Barnes}, \citet{1995Hernquist} and others. Observational results confirm that mergers accompany nuclear activity in some sources \citep{2001Canalizo, 2011Ellison}, but there is growing evidence that mergers are not the main kpc-scale driving mechanism of AGN activity, at least at moderate luminosity \citep{2011Cisternas, 2011Ellison, 2012Kocevski, 2014Villforth}. The driver of kpc-scale galactic gas inflow likely depends on AGN luminosity and redshift \citep{2009Hopkins, 2012Treister, 2014Comerford}. Transient interactions between galaxies, or non-axisymmetric gravitational instabilities in isolated disk galaxies \citep{2000Cavaliere, 2010Hopkins, 2015Gatti}, can also drive gas inflows. At high redshift, torques may be supplied by massive star-forming regions in gas-rich galaxies; such ``clumps'' are most frequently seen at redshift of $\sim 2$ \citep{2016Shibuya} and models suggest that they sink in the gravitational potential of their host, driving gas inflow \citep{1998Noguchi, 2007Bournaud, 2008Genzel}.
 
 %\textcolor{red}{\bf The kpc-scale ang. mom. extraction mechanisms fail at $r\sim1$ kpc. Possible solution: ``stuff within stuff'' model.}
 
 However, torques induced by large-scale structures cannot efficiently extract angular momentum at distances $< 1$~kpc from the galactic center \citep{2003Goodman}. Viscosity-related effects are not efficient enough beyond the last parsec from the black hole \citep{1989ShlosmanBegelman, 2003Goodman}. Gravitational torques may again be a viable solution. The ``bars within bars'' model, originally proposed by \citet{1989Shlosman}, assumes the presence of a series of embedded bars, stretching all the way from kpc-scales to the central SMBH accretion structures. These structures are expected to gradually remove angular momentum from the gas, letting it reach the galactic center sufficiently fast to explain AGN activity. The ``bars within bars'' mechanism, later revised to the ``non-axisymmetric features all the way down'' model (also referred to as the ``stuff within stuff'' model) to account for non-axisymmetries other than bars \citep{2010Hopkins, 2011Hopkins}, has been found to arise in close to self-consistent nested zoom-in simulations of \citet{2010Hopkins, 2011Hopkins} and tested in several high-resolution numerical studies \citep{2007Escala, 2016AnglesAlcazar}. Moreover, central $\sim 100$~pc scale dust structures with various morphologies have been observed in a number of active and inactive galaxies \citep{1997Jungwiert, 1999Regan, 2000Marquez, 2003Martini} and their morphologies confirm the viability of gravitational instabilities as the main gas inflow mechanism at $\sim 100$~pc scales \citep{2000Maiolino, 2009Davies, 2009Haan, 2014Combes, 2014Davies}.
 
 %\textcolor{red}{\bf Our plan: search for misalignments!}
 
 If the ``non-axisymmetric features all the way down'' model is true, how do its features manifest in observations? One general feature is theoretically predicted by \citet{2012Hopkins}. They report that the non-axisymmetric structures are expected to progressively misalign from the disk of their host galaxies as they reach further into the galaxy. Here, we search directly for these structures on $\sim 100$~pc scales using \emph{HST} data, for a special sample of AGN where we know the orientation of the accretion disk precisely.
 
 %\textcolor{red}{\bf Why do we need H2O megamasers?}
  
 An especially precise measurement of the orientation of the central accretion flow in galactic nuclei is possible with observations of H$_2$O megamasers (see \citealt{2013Greene} and a review by \citealt{2005Lo}). The maser emission in these systems originates in a ring of material illuminated by the AGN at a distance of~$\sim 0.1-0.5$~pc from the black hole. The ring is located in the viscous-torque-dominated region of gas inflow and is thus expected to align with the accretion disk around the black hole. These megamaser disks are only detected when they are close to edge-on, since in that orientation the optical depth for maser action is maximized. Mapping with VLBI at sub-pc resolution (e.g., \citealt{1990Greenhill, 2011Kuo}, see also \citealt{2005Lo}) provides a very precise three-dimensional orientation for the accretion disk, a great advantage over other methods. Combined with HST/WFC3 images of the host galaxy, these properties make them excellent targets for investigation of how the orientation of nuclear non-axisymmetries correlates with the orientation of SMBH's accretion disk. Such a comparison, along with morphological characterization of identified nuclear structures, is the main goal of this work.
 
 %\textcolor{red}{\bf Paper layout}
 
 The paper is organized as follows. In Sect.~\ref{sect:obs} we describe the instruments and technical details of the observations, as well as initial data reduction leading to the results presented in the following sections. In Sect.~\ref{sect:data_analysis}, the methods of identification and classification of the nuclear regions are presented. Calculations related to the orientation of galactic structures based on optical images are described in Sect.~\ref{sect:orient}. Our results concerning the morphologies of nuclear regions are given in Sect.~\ref{sect:morphology} and the relative orientation of various components within our galaxies is presented in Sect.~\ref{sect:results_angmom}. We discuss and summarize our findings in Sect.~\ref{sect:discussion}. In Appendix~\ref{sect:3rdpageplots} we present detailed information on our analysis of each of the 9 new megamaser galaxies. In Appendix~\ref{sect:classVsDist} we consider how the morphological classification of the nuclear regions in our sample depends on galaxy distance, scale of the region, and available resolution.
 
 \section{Observations and data reduction}\label{sect:obs}
 
 There are 34 known megamaser disk galaxies \citep{2015Pesce}. We focus here on a subset of 18 megamaser disk galaxies with reliable BH mass measurements from Keplerian fitting to the maser dynamics (\citealt{2011Kuo, 2016Gao, 2016Gao-b} and W.~Zhao~et~al.~2017, in prep.). The list of our sources is given in Table~\ref{tab:orient}. In general, the galaxies are early-type spiral galaxies (e.g., \citealt{2010Greene}) and we have studied the detailed morphological structure of roughly half of the galaxies in \cite{2016Laesker} using the \emph{Hubble Space Telescope} (HST).
 
 Each target was observed in two orbits with \emph{HST} between Dec 1$^{\rm st}$ 2014 and Aug 29$^{\rm th}$ 2015. We obtained F336W, F438W, F814W, F110W, and F160W (roughly UBIJH) images of each galaxy with integration times of 1320, 430, 2140, 150, and 420~s, respectively. In the optical, we use a three-point dither pattern for cosmic-ray removal, and in the NIR we use the 4-point dither pattern. We use the default output of the \texttt{MultiDrizzle} pipeline, which performs cosmic-ray rejection and optimally combines the images.
 
 \section{Data analysis}\label{sect:data_analysis}
 
 The goal of this paper is to characterize the innermost structures in galaxies with maser disks. Here we describe the classification of these structures and the process of deriving their orientation from \emph{HST} data.
 
 \subsection{Ellipse fitting}
 
 In order to support the identification of nuclear structures in our sample of galaxies, we used the algorithm of \cite{1987Jedrzejewski} to fit ellipses to the galaxy isophotes, implemented as the \texttt{IRAF} script \texttt{ellipse}\footnote{\texttt{ellipse} is included in \texttt{STSDAS} (version 3.17) available as a package for \texttt{IRAF} (version 2.16.1 used).}.
 
 The initial analysis was performed using \texttt{IRAF}. First, the foreground stars and background galaxies were manually masked (in \texttt{DS9}\footnote{SAOImage DS9, \url{http://ds9.si.edu}.}). The centers of the galaxies in NIR- and UVIS-band filters were found using the task \texttt{imexamine} and fed to an \texttt{ellipse} parameter file as initial ellipse center positions. The ellipse centers were then further refined by \texttt{ellipse}. In some cases (all filters for Mrk~1029, F336W for ESO~558 and J0437+2345, F438W for UGC~3193, as well as F336W and F438W for NGC~5765b), the ``object locator's k-sigma threshold'' was also lowered from the default value of $1.0$ to $0.5$ in order for the algorithm to correctly identify the galactic center.
 
 For each of our targets, the \texttt{ellipse} run conducted as above resulted in a list of elliptical fits to image isophotes for each of the filters -- we use the parameters of those fits in further analysis.
 
 \subsection{Structure maps} 

 As a second method to characterize nuclear morphology, we use structure maps to remove large-scale smooth galaxy components and highlight the small-scale features (such as dust lanes). 
 The concept of structure maps was introduced by \citet{2002Pogge}. The technique is designed to remove low-frequency (smooth) features of the map, and highlight high-frequency features around the scale of the PSF. The method is closely related to Richardson-Lucy deconvolution \citep{1972Richardson, 1974Lucy}. It is also similar in spirit to unsharp masking, but with structure maps the convolution is done with the PSF itself rather than a boxcar.
  Mathematically, a structure image is given by eq.~(1) of \cite{2002Pogge}:
\beq S = \left[ \frac{I}{I\otimes P}\right]\otimes P^T \textrm{,} \label{eq:struct} \eeq 
 where $S$ is the structure image pixel matrix, $I$ is the original image, $P$ is the point-spread function (PSF), $P^T$ is the transposed PSF and $\otimes$ denotes convolution. Structure maps emphasize high frequency features that are nearly unresolved in the original image.

 %\textcolor{red}{As argued by \citet{2002Pogge}, structure maps are, in principle, analogous to the '$2^{\textrm{nd}}$-order correction image' of the Richardson-Lucy image reconstruction method \citep{1972Richardson, 1974Lucy}.}
 %\textcolor{red}{which are not represented in the 1$^{\textrm{st}}$ order, smooth reconstruction image.}
 
 %We have produced structure maps using the F814W images, proceeding as follows. We extracted a PSF from each of the images by selecting a $49 \times 49$~px box centered on an isolated point-like source (star). The PSF image was recentered and over-sampled, i.e. each of its pixels was transformed into a $4\times 4$ matrix using bicubic interpolation\footnote{as implemented in \texttt{Python} 2.7 package \texttt{cv2} (version 3.1.0).}. The over-resolved PSF was then centered on its brightest pixel and resolution was returned to initial using bicubic interpolation, resulting in a $45\times 45$ pixel PSF.
 %\textcolor{blue}{do not think we need this much detail for recentering. Just say sky-subtracted and recentered.} Next, sky-subtraction was performed on each re-centered PSF. The sky level was approximated by averaging count rates in  $4\times4$~px regions in each of the corners of the PSF and the resulting average count rate was subtracted from each of the pixels of the PSF image.
 
 In our analysis we used structure maps to highlight dust features in the galaxies of our sample. Two filters in our data set, F438W and F814W, are suitably sensitive to dust to derive the structure maps from them. F438W provides higher spatial resolution and is more sensitive to dust. However, in our data the F814W filter has significantly better signal-to-noise ratio than F438W. In our analysis we have therefore used the structure maps based on the F814W images.
 
 The structure map derivation proceeded as follows. We extracted a sky-subtracted and re-centered PSF from each of the images. A lower cutoff $n_0$ for the count rate per pixel of $0.00001$ was set and all the pixels on both the original F814W image and the PSF with count rates $n<n_0$ were assigned count rates $n_0$ in order to avoid division by zero and negative values. Each of the F814W images was then convolved with its respective PSF obtained in previous steps\footnote{For Mrk~1210 and UGC6093 the PSFs obtained using point-sources in their F814W images were strongly asymmetric, resulting in dipole-like artifacts visible on the structure maps. Therefore we decided to use a more stable PSF from NGC~5495 in calculation of the structure maps in these two cases.}. Finally, a copy of the original image was divided by the result of the previous step and the resulting image was convolved with a transposed PSF template. The structure maps for all the new sources can be found in Figs.~\ref{fig:3rdPage1} and\ \ref{fig:3rdPage_ESO558}~--~\ref{fig:3rdPage_UGC6093}.
 
 The re-analysis of the sources of \citet{2013Greene} was performed analogously.
 
 \subsection{Identifying nuclear structures}
 \label{sect:identNuclStruct}
 
 %\subsection{Definitions}\label{sect:definitions}
 
 We consider a range of scales in this work: the $0.1$--$0.5$~pc scale of the megamaser disk, the nuclear scales of $100$--$500$~pc where we seek kinematically cold or flattened structures (``nuclear regions'') along with spiral features (``nuclear spirals'') and the galaxy-wide disk on kpc scales.
 
 %Throughout this work, we refer to several regions of interest in each galaxy. These are:
% \begin{itemize}
%     \item the megamaser disk -- a ring of material at the distance $0.1$~--~$0.5$~pc from the galactic center, at which the observed H$_2$O maser emission originates; this material belongs to the viscosity-dominated region of the SMBH accretion flow and is expected to align with the central accretion disk (or be part of it);
%     \item the nuclear region -- a ring of material $50$~--~$600$~pc in radius, corresponding to the smallest resolved structure of the \emph{HST} images presented in this work; the size / radius of the nuclear region corresponds to the radius of the outer edge of such innermost resolved structure;
%     \item nuclear spiral -- spiral dust structure accompanying some of the nuclear regions;
     %\item \textcolor{red}{intermediate-scale regions -- larger features of interest visible in the galaxy image;}
%     \item the kpc-scale galaxy -- range of radii from the galactic center at which isophotes in \emph{HST} images correspond to constant-radius lines in the galactic disk.
% \end{itemize}

 Many galaxies are known to have nuclear disks on $10$-$100$~pc scales (e.g., \citealt{2014Combes, 2014Garcia-Burillo}) and we resolve the $100$~pc scales in $10$ of our $18$ galaxies. With the theoretical resolution limits of the F110W ($\sim1153.4$~nm), F160W ($\sim1536.9$~nm), and F814W ($\sim802.4$nm) filters being $0.12$, $0.16$ and~$0.08$~arcsec, respectively, we identify nuclear structures at least $0.3$~arcsec in size (which limits the visibility of $100$~pc structures to $\sim 70$~Mpc; see Table~\ref{tab:orient} for distances to the galaxies in our sample). The outer radii of our nuclear regions range between $0.3$ and $2.3$~arcsec, with a median of $0.6$~arcsec. In most cases we operate at the very limits of what can be robustly resolved and identified. However, we sometimes select larger structures that have a clearer interpretation to be able to analyse morphology and~/~or be better equipped to extract the orientation of the nuclear regions.
 
 To ensure that the lower limit of the nuclear region's angular size of $0.3$~arcsec is sufficient, we have re-derived all the results presented in Sections~\ref{sect:morphology} and~\ref{sect:results_angmom} adopting a more restrictive limit on the angular size of a nuclear region of $0.5$~arcsec (i.e., excluding 8 galaxies that host nuclear regions with radii $<0.5$~arcsec from the analysis). This corresponded to removing most of the galaxies beyond $100$~Mpc from the sample. Our conclusions (see Sections~\ref{sect:morphology} and~\ref{sect:results_angmom}) remained mostly unchanged -- for specific results of this trial and their discussion see Appendix~\ref{sect:morph_scale}.
 
 %It is also worth noting that while we do classify nuclear spirals in our sample, we do not treat them as nuclear regions themselves, but rather as accompanying features. This is because isophotes in spiral regions do not correspond to constant-radius lines in the local nuclear disk of the galaxy (indeed, PA is expected to rotate tracking the pattern) and hence, we would not be able to extract information about orientation of the part of the nuclear disk where the nuclear spiral is present. Instead, we choose nuclear rings or other features as the nuclear regions and classify the nuclear spirals separately (see, e.g.,NGC5875b, Fig.~\ref{fig:3rdPage3}).
 
 With \texttt{ellipse} fits and structure maps in hand, we proceed to identify nuclear structures in our sample galaxies. While we took into account all the available filters, we concentrated our efforts on F110W (or F160W) and F814W (the deepest image in the UVIS band); with F110W tracking starlight and F814W$-$F110W interstellar dust. For each of our galaxies we have chosen a set of ellipses that we associate with the large-scale galaxy, assuming that the isophotes on large scales are fit by an axisymmetric disk. These define the ``kpc-scale'' galaxy to which we refer in the following sections. We also identify a set of ellipses associated with the outer edge of a nuclear structure of size $\sim50-600$~pc. We carefully pick those isophotes to correspond to changes in PA and ellipticity profiles, so that they correspond to a physical feature in the galactic nucleus (for details on how the structures are identified see Sect.~\ref{sect:classification}).
 
 %\textcolor{red}{By default, \texttt{ellipse} returns the ellipticity. For minor and major semi-axes $b$ and $a$, the ellipticity is given by $\epsilon = 1-\frac{b}{a}$. Throughout this paper we use eccentricity instead ($e = \sqrt{1-(1-\epsilon)^2}$).}
 
 \subsection{Classification of nuclear structure}\label{sect:classification}
 
 To discuss the morphology of nuclear structures, we classify them in two ways. First, we assign a class to the region itself, according to the key:
 \begin{itemize}
  \item D -- disk,
  \item R -- ring,
  \item Bu -- bulge,
  \item B -- bar,
  \item Ch -- no discernible morphology, chaotic dust structure.
 \end{itemize}
 An additional ``?'' sign marks class assignment as unsure.
 
 Nuclear spirals are almost ubiquitously found in late-type spiral galaxies (e.g., \citealt{2002Pogge, 2003Martini}). We therefore add a classification of potential spiral dust structure surrounding our set of nuclear isophotes, which we append to the nuclear region classification after a ``$+$'' sign:
 \begin{itemize}
  \item N -- no surrounding spiral dust structure;
  \item Sx -- spiral structure visible, where x denotes its type (``gd'' -- two-arm grand-design, ``tw'' -- tightly wound or flocculent).
 \end{itemize}
 As an example, class D/R+Stw is assigned to a galaxy with a central disk or ring with a tightly-wound spiral structure -- as in the case of~Mrk~1210, see Fig.~\ref{fig:3rdPage_Mrk1210} and Table~\ref{tab:orient}.
 
 The outputs from \texttt{ellipse} aid our identification of nuclear structures in the following manner (cf. \citealt{2013Greene}):
 \begin{itemize}
  \item bars are characterized by a region of constant position angle (PA), ellipticity ($\epsilon$) decreasing inwards, and rapid $\epsilon$ and PA changes at their outer edge \citep{2002Maciejewski, 2003Erwin} -- see, e.g., the $\sim 4''$-scale bar in UGC~6093, Fig.~\ref{fig:3rdPage_UGC6093};
  \item spiral structure is identified by smoothly rotating position angle with $\epsilon$ constant or changing \citep{2003Martini} -- see the spiral structure outside $5''$ in UGC~6093, Fig.~\ref{fig:3rdPage_UGC6093};
  \item disks can be recognized by relatively constant PA and significant ellipticity -- see the nuclear disk at $\sim 0.6''$ in Mrk~1210, Fig.~\ref{fig:3rdPage_Mrk1210};
  \item rings exhibit features similar to disks, but are distinguished by discontinuities in PA at their edges, as well as ``bumps'' in the surface brightness profiles \citep{1986Buta} -- see $\sim2''$ nuclear ring of NGC~5728, Fig.~\ref{fig:3rdPage_NGC5728};
  \item while the \texttt{ellipse} results should show bulges as round (i.e., with low ellipticity), with constant PA, and a surface brightness profile that is steadily rising towards the center (see, e.g., the $\sim3''$-scale bulge in UGC~3193, Fig.~\ref{fig:3rdPage_UGC3193}), confident classification of bulges requires detailed 2D modelling of a galaxy (see, e.g., \citealt{2016Laesker}); we do not attempt such decomposition here and, therefore, nuclear regions classified as bulges in this work should be treated as tentative.
 \end{itemize}
 
 These considerations are only one part of our analysis. We also consider color and structure maps. For example, flocculent or tightly wound nuclear spirals are best detectable with structure maps, which improve the visibility of any PSF-scale structure, regardless of its symmetry, while they would fall below the spatial resolution of \texttt{ellipse} profiles. In turn, if narrow line regions (NLRs) are present in a galaxy image, they are best distinguished from dust structures by color maps, on which they appear very blue. The NLR emission may hide any dust components in the structure map, but the contamination due to the NLR can be visible in a color image (see the conical structure in NGC~5728, Fig.~\ref{fig:3rdPage_NGC5728}, which we interpret following \citealt{1988Schommer} and \citealt{1993Wilson} as an AGN ionization cone). Finally, the F110W data traces stellar light, and allows us to distinguish features corresponding to the stellar component of galactic nuclei, enabling us to verify whether a circular feature visible in the galaxy image may correspond to a bulge.
 
 In some parts of our analysis it is beneficial to divide the nuclear regions in our sample into smaller and larger regions. The radius separating these two groups, $r_b=200$~pc, provides an equal number of objects in each size bin: there are 9 small nuclear regions ($r<200$~pc) and 9 large nuclear regions ($r\ge200$~pc); see Table~\ref{tab:orient}. We admit that the $200$~pc boundary is somewhat arbitrary, but small changes in this value that roughly keep sample sizes similar do not yield different results of our analysis. As noted in Sect.~\ref{sect:identNuclStruct}, all of our nuclear structures were identified at sizes of at least $0.3$~arcsec. While re-deriving all the results presented in Sections~\ref{sect:morphology} and~\ref{sect:results_angmom} with a more restrictive limit on angular size of a nuclear region of $0.5$~arcsec (see Sect.~\ref{sect:identNuclStruct} and Appendix~\ref{sect:morph_scale}), we also made sure that the determination of orientation for small ($r_b<200$~pc) nuclear regions is not affected by resolution effects. In the restricted sample the differences between large and small nuclear regions are still apparent and the sample is still evenly divided at $200$~pc -- for specific results see Appendix~\ref{sect:morph_scale}.

 \subsection{Angular momentum orientation from \texttt{ellipse} fits}\label{sect:orient}
 
 If a nuclear region is disk- or ring-like, the orientation of its angular momentum in space is easily recovered from the \texttt{ellipse} fits. For each source we calculate the positions of the angular momenta of the galaxy as a whole and the nuclear region, utilizing the groups of ellipses described in Sect.~\ref{sect:identNuclStruct}. We take the average position angle (PA) and eccentricity ($e = \sqrt{1-(1-\epsilon)^2}$) within the region and calculate $i=\arccos{\left(\sqrt{1-e_{\rm avg}^2}\right)}$. Note that the PA we quote is for the angular momentum vector and is thus aligned with the minor axis of the projected disk. For instance, $\textrm{PA}=30\degree$ describes a disk with its projected angular momentum pointing $30\degree$ from North towards East on the sky and the image of the nuclear region in this case appears elongated along the $\textrm{PA} = 120\degree$ direction with the blue-shifted edge in the direction of $\textrm{PA} = 120\degree$ and the red-shifted edge in the direction of $\textrm{PA} = 300\degree$. Four possible three-dimensional angular momentum directions are allowed by $\textrm{PA}_{\rm avg}$ and $i$. The position angle can be $\textrm{PA}_{\rm avg}$ or $\textrm{PA}_{\rm avg}+180\degree$ and an inclination of either $i$ or $180\degree-i$ is allowed (where $i=0$ corresponds to the angular momentum of the disk pointing towards the Earth). We are {\it assuming} that the nuclear region is a disk (treated as infinitely thin) to make this assignment. We quantify the accuracy of such an assumption by assigning a class to each nuclear region (see Sect.~\ref{sect:classification}). The nuclear bulges are ignored in our analysis of nuclear regions' orientations as their orientation cannot be established using the method described above. We do, however, consider the 3 nuclear regions with chaotic dust structures in our sample to correspond to flattened structures, and take their orientations into account.
 
 There are important biases associated with our estimation of 3D positions of angular momenta, especially for nuclear regions. If a nuclear region is edge-on, \texttt{ellipse} will still fit a finite-width ellipse to its isophotes due to vertical structure in the disk and finite PSF, giving $i\neq0$ (the floor appears to be $i_{\rm min}\sim20\degree$, see Table~\ref{tab:orient}). We also avoid face-on nuclear regions due to random structure in the plane of the disk distorting \texttt{ellipse} fits and potential misclassification of face-on disks as bulges.
 
 In the case of the set of ellipses associated with the kpc-scale disk, additional information can be used to further constrain the orientation of the angular momentum. All of the galaxies in our sample are spirals. If we assume that the spiral arms are trailing\footnotetext{In the case of NGC~5728, which has two sets of large-scale spiral arms wound in opposite directions, we assumed the inner set to be trailing for consistency with other sources (where the spiral structure is often only visible to a limited distance from the center).}, as is observed in most spiral galaxies (see \citealt{BinneyTremaine} and references therein), this fixes the inclination and leaves only two possible orientations. In some cases, the rotation curves of the galaxies are also available, leaving only one angular momentum orientation allowed by the data.
 
 For the galaxies without rotation curves available, we have used the relative prominence of dust lanes in the galaxy to constrain the orientation, a method originally suggested by \cite{1929Hubble} and used by, e.g., \cite{1985Sharp, 2008Vaisanen}. The dust lanes of the part of the galaxy in front of its nucleus as seen by the observer are expected to be more pronounced than those behind it due to their being back-lit by stronger galactic emission closer to the nucleus. The only galaxy where a rotation curve is not available and dust lanes prominence method does not yield a reliable orientation (due to the galaxy being almost face-on) is NGC~5495, where we have kept both possible position angles of the angular momentum in the analysis. Table~\ref{tab:orient} gives all the resulting PA and inclination values for each of the galaxies.
 
 In our investigations, we also use results and expand the analysis of \cite{2013Greene} in order to derive the statistics of the total sample of $18$ sources. The data related to these galaxies are included in the summary in Table~\ref{tab:orient}.
  
 \section{Results: Morphology of nuclear regions}\label{sect:morphology}
 
\begin{figure*}
  \centering
  \makebox[\textwidth]{\includegraphics{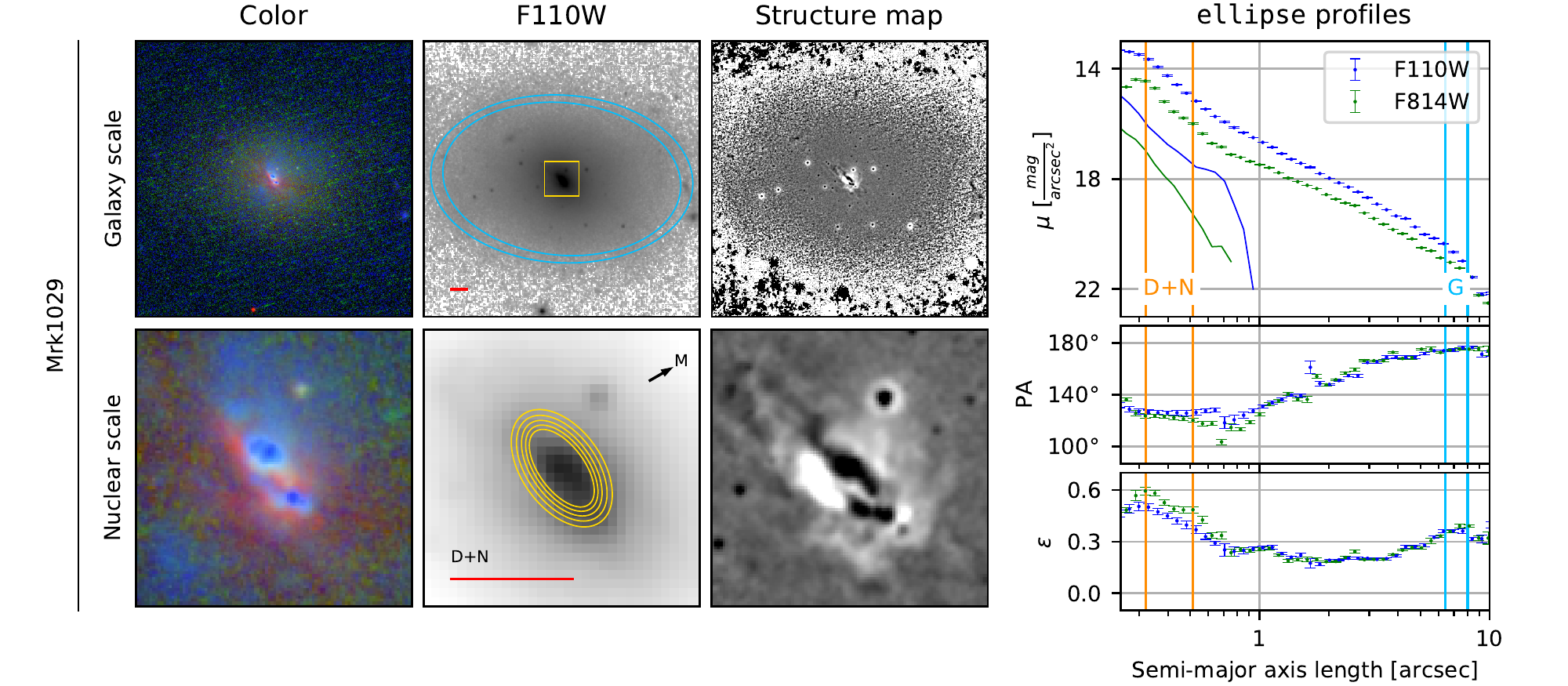}}
  \makebox[\textwidth]{\includegraphics{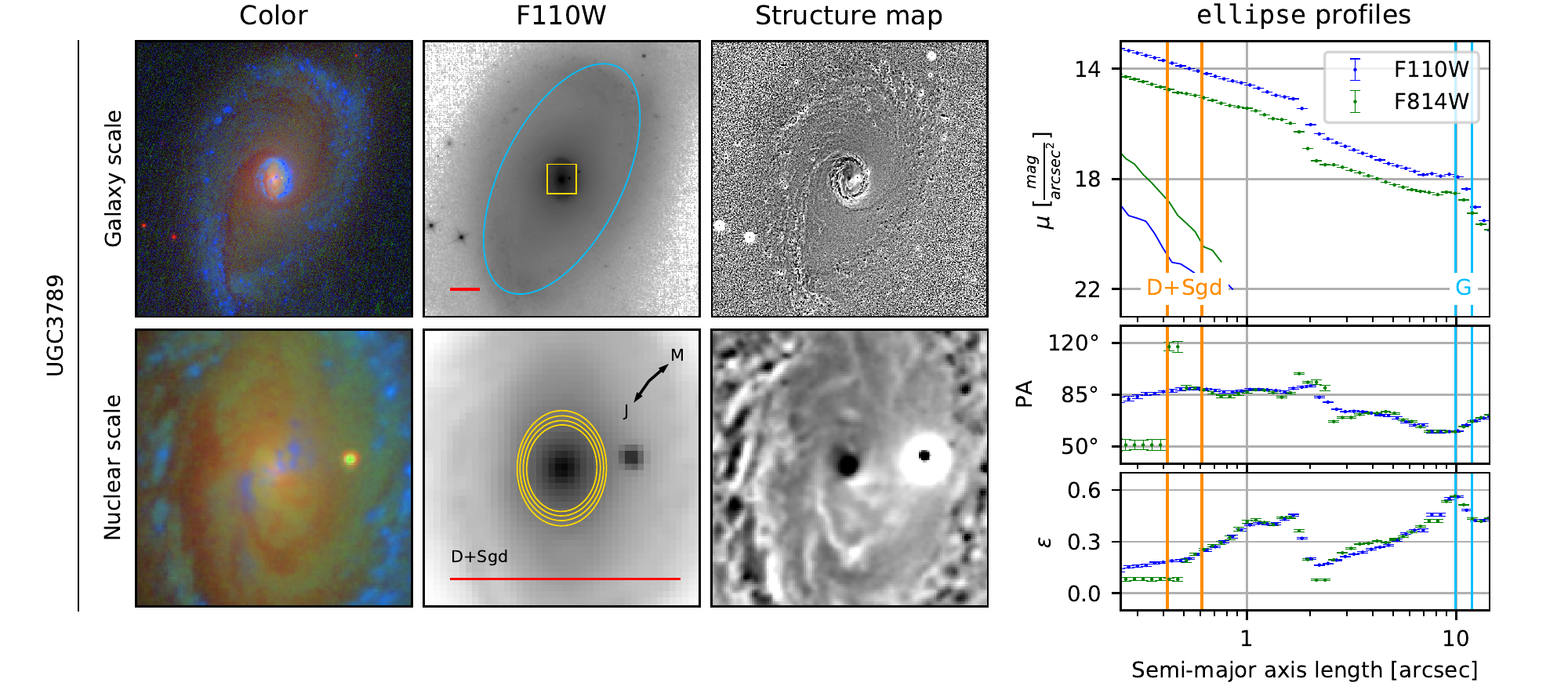}}
  \caption{Images and \texttt{ellipse} fits of two of the galaxies in our sample (for analogous figures for the remaining eight of the nine new galaxies see Appendix~\ref{sect:3rdpageplots}; UGC~3789 was taken from the \citealt{2013Greene} sample). In all images North is up and East is left. Each row shows images of the same field of view (top in each panel -- galaxy scales, bottom -- nuclear scales). Column 1 (from left to right): false-color images (blue -- F335W, green -- F438W, and red -- F814W). Column 2: F110W image (color -- logarithmic scale for count rate). Column 3: structure maps constructed from the F814W image (logarithmic scale). In the F110W image in the top row (second column), ellipses following the galaxy-wide orientation are marked in blue and a yellow rectangle shows the region presented in the bottom row. In the F110W image in the bottom row, ellipses tracing the nuclear region are marked in yellow. The position angle of the maser disk angular momentum vector, perpendicular to the line of nodes of the masing disk, and the PA of the jet (if known) are shown in the upper right corner of this image as black arrows marked with ``M'' and ``J'', respectively. Jet orientation references for all sources where such data were available (see Appendix~\ref{sect:3rdpageplots}): \citet{1988Schommer, 1998Falcke, 2001Schmitt, 2009Mundell, 2010Xanthopoulos, 2012Yamauchi, 2013Sun} and the FIRST survey, \cite{FIRST}. The red bar at the bottom of each F110W image is $1$~kpc in projected distance. Nuclear class (see Sect.~\ref{sect:classification}) is noted above the red bar on the nuclear-scale image. Column 4: Surface brightness ($\mu$, blue and green circles with horizontal bars indicating the angular range), position angle (PA) and ellipticity ($\epsilon$) profiles from \texttt{ellipse} for the F110W (blue) and F814W (green) images. Note that eccentricity $e=\sqrt{1-(1-\epsilon)^2}$. Vertical blue and orange lines limit the ranges of ellipse major axes used to extract the orientation of a galaxy as a whole and the nuclear region, respectively. The type of each structure is indicated. Blue and green solid lines on the surface brightness plots show \texttt{ellipse} fits to point sources in F110W and F814W images, respectively, approximating the point-spread function (PSF). The PSF profiles have been artificially scaled in brightness to optimize their visibility on the plots.}
  \label{fig:3rdPage1}
 \end{figure*}

 Figures~\ref{fig:3rdPage1} and\ \ref{fig:3rdPage_ESO558}~--~\ref{fig:3rdPage_UGC6093} show the ellipses corresponding to the large-scale galaxy and the nuclear region.
 
 When it comes to the nuclear regions, our sample of 18 galaxies contains $12$ disky structures, $4$ of which are rings, $2$ are disk/ring structures, and $6$ do not exhibit any additional morphology. We also identify $3$ bulges and $3$ chaotic dust structures. There are $3$ sources with grand-design nuclear spirals and $5$ with tightly-wound ones. Two of the former belong to galaxies with a large-scale bar (the exception being UGC~3789), while all of the latter belong to non-barred galaxies. Ten galaxies do not show any nuclear spirals associated with the identified regions.
 
 \begin{table*}
 \centering
 \begin{threeparttable}
  \caption{Orientation of the masing disk, galaxy and nuclear region for each of the galaxies.}
  \renewcommand*{\arraystretch}{1.2}
  \begin{tabular}{|c|l|c|ccc|ccc|ccccc|}
  \hline
  & \multirow{2}{*}{Object} & $D_A$ & \multicolumn{3}{c|}{Maser} & \multicolumn{3}{c|}{Galaxy} & \multicolumn{5}{c|}{Nuclear region} \\
  \cline{4-14}
  & & [Mpc] & $i$ & PA & ref. & $e$ & $i$ & PA & Class & $r$ [pc] & $e$ & $i$ & PA \\
  (1) & \multicolumn{1}{c|}{(2)} & (3) & (4) & (5) & (6) & (7) & (8) & (9) & (10) & (11) & (12) & (13) & (14) \\
  \hline
  \multirow{9}{*}[-0em]{\begin{sideways}This work\end{sideways}}  & ESO558$-$G009 & 109 & 90 & 350 & G16p$\dagger$ & 0.97 & $104$ & $277^{\rm d}$ & Ch+N & 210 & 0.87 & 61 & 99 \\
 & J0437+2456 & 70 & 90 & 115 & G16p$\dagger$ & 0.82 & $125$ & $126^{\rm d}$ & Bu+N & 120 & 0.77 & 51 & 131 \\
 & Mrk1029 & 117 & 90 & 300 & G16p$\dagger$ & 0.77 & $130$ & $176^{\rm d}$ & D+N & 220 & 0.83 & 57 & 126 \\
 & Mrk1210 & 58 & 101 & 333 & Z17p & 0.33 & $19$ & $36^{\rm r}$ & D/R+Stw & 170 & 0.45 & 26 & 119 \\
 & NGC5495 & 93 & 90 & 270 & G16p$\dagger$ & 0.63 & $141$ & $119, 299$ & D/R+Sgd & 170 & 0.49 & 29 & 57 \\
 & NGC5728 & 41 & 90 & 329 & K17p$\dagger$ & 0.91 & $114$ & $302^{\rm r}$ & R+Stw & 460 & 0.36 & 21 & 121 \\
 & NGC5765b & 126 & 95 & 237 & G16 & 0.67 & $42$ & $337^{\rm r}$ & R+Stw & 450 & 0.45 & 27 & 67 \\
 & UGC3193 & 60 & 90 & 60 & W17p$\dagger$ & 0.95 & $72$ & $80^{\rm d}$ & D?+N & 220 & 0.90 & 65 & 79 \\
 & UGC6093 & 147 & 94 & 70 & Z17p & 0.47 & $152$ & $0^{\rm r}$ & Bu?+N & 150 & 0.26 & 15 & 92 \\

  \hline
  \multirow{8}{*}[-0.8em]{\begin{sideways}\cite{2013Greene}\end{sideways}}  & IC2560 & 44 & 90 & 44 & Y12 & 0.88 & $62$ & $133^{\rm r}$ & D+Stw & 100 & 0.74 & 48 & 136 \\
 & NGC1194 & 52 & 85 & 67 & K11 & 0.84 & $57$ & $50^{\rm r}$ & Ch+N & 120 & 0.63 & 39 & 68 \\
 & NGC2273 & 26 & 84 & 63 & K11 & 0.68 & $43$ & $176^{\rm r}$ & R+Stw & 150 & 0.86 & 59 & 131 \\
 & NGC2960 & 71 & 89 & 139 & K11 & 0.64 & $140$ & $314^{\rm r}$ & D+N & 260 & 0.83 & 56 & 139 \\
 & NGC3393 & 55 & 90 & 56 & K08$\dagger$ & 0.72 & $46$ & $252^{\rm r}$ & R+Sgd & 270 & 0.68 & 43 & 58 \\
 & NGC4388 & 19 & 90 & 17 & K11$\dagger$ & 0.98 & $78$ & $0$ & D+N & 100 & 1.00 & 89 & 345 \\
 & NGC6264 & 134 & 90 & 185 & K11 & 0.79 & $52$ & $290^{\rm d}$ & Ch+N & 290 & 0.68 & 43 & 123 \\
 & NGC6323 & 103 & 89 & 280 & K11 & 0.94 & $70$ & $85^{\rm r}$ & Bu?+N & 200 & 0.62 & 38 & 95 \\
 & UGC3789 & 45 & 89 & 311 & K11 & 0.87 & $120$ & $243^{\rm r}$ & D+Sgd & 100 & 0.62 & 39 & 88 \\

  \hline
  \end{tabular}
  \begin{tablenotes}\footnotesize
   \item \textbf{Notes:} Column 1 -- data sample. Column 2 -- galaxy designation. Column 3 -- angular diameter distance from NED (\url{http://ned.ipac.caltech.edu/}; with its default cosmology: $H_0 = 73$~km~s$^{-1}$~Mpc$^{-1}$, $\Omega_m = 0.27$, $\Omega_{\Lambda}=0.73$) or using data from \citet{2016Gao} in the case of NGC~5765b. Columns 4-6 -- inclination ($i$) and position angle (PA) of the maser disk as reported in reference in column~6. Hereafter $i < 90\degree$ corresponds to a feature whose angular momentum is directed towards Earth and $i > 90\degree$ -- one whose angular momentum is directed away from it. References with $\dagger$ do not constrain inclination and $i=90\degree$ is assumed there. Columns 7-9 -- large-scale galaxy orientation. $e$ denotes the average eccentricity of \texttt{ellipse} fits identified as following the orientation of the galaxy as a whole (see Sect.~\ref{sect:identNuclStruct}). Note that eccentricity $e$ is given by ellipticity $\epsilon$ through $e = \sqrt{1-(1-\epsilon)^2}$. $i$ and PA are the inclination and position angle resulting from $e$, PA$_{\rm avg}$, rotation curves of the galaxies (superscript ``r''; references: \citealt{1988Schommer, 2000Cooke, 2003Schulz, 2013Fischer, 2015Bosch}) or relative prominence of their dust lanes (superscript ``d''; see Sect.~\ref{sect:orient}), and the assumption that the spiral arms of all the galaxies are trailing. For NGC~4388, instead of \texttt{ellipse} fits, kinematic arguments from \citet{2014Greene} are used to constrain the nuclear disk and galaxy orientation. Columns 10-14 -- nuclear region orientation based on the smallest scales at which such measurement is possible with our data. Column 10 -- morphological class (see Sect.~\ref{sect:classification}). Classes of the nuclear region: D -- disk, R -- ring, Bu -- bulge, B -- bar, Ch -- no discernible morphology. Classes of the nuclear spiral: N -- no surrounding spiral dust structure, Sx -- spiral structure visible, where x denotes its type (``gd'' -- two-arm grand-design, ``tw'' -- tightly wound or flocculent). Column 11 -- outer physical radius of the nuclear region (disk assumed). $e$ -- average eccentricity of ellipse fits identified as following the orientation of the nuclear region. $i$ -- smaller of the two possible inclinations allowed by $e$. PA -- smaller of the two possible position angles allowed by the data.
   \item \textbf{References:} G16 -- \cite{2016Gao}; G16p -- \cite{2016Gao-b}; K08 -- \cite{2008Kondratko}; K11 -- \cite{2011Kuo}; K17p -- Kuo et al. (2017), in prep.; W17p -- Wagner et al. (2017), in prep.; Y12 -- \cite{2012Yamauchi}; Z17p -- Zhao et al. (2017), in prep.
  \end{tablenotes}
  \label{tab:orient}
  \end{threeparttable}
 \end{table*}
 
 \subsection{Comparison with previous work}\label{sect:morph_comparison}
 
 We see a significant amount of structure in the galactic nuclei, consistent with the predictions of the ``non-axisymmetric features all the way down'' model and previous studies of nuclear dust structures in active galaxies \citep{2002Xilouris, 2007SimoesLopes, 2013Hicks}. Throughout, we assume that the megamaser disk galaxies are a fair sample of spiral galaxies, but in principle we do not know this, and must bear in mind the caveat that the masers may be a special subset of all spiral galaxies \citep[see, e.g.,][]{2016GreeneSMBHmass}.
 
 \cite{2003Martini} performed a detailed characterization of nuclear spirals in their matched galaxy samples ($28$ pairs of active/inactive galaxies and $18$ pairs of barred/unbarred ones). As we conduct a similar morphological classification here, it is worthwhile to compare our results with theirs. \cite{2003Martini} report the presence of nuclear spirals for $75\%$ of their 28 active galaxies. In our sample this fraction is considerably lower: 8 out of 18 galaxies, i.e., $44\%$. The results are, however, marginally consistent with each other if nominal uncertainties in each sample of $\sim 16\%$ are taken into account.
 
 %The uncertainty of the fraction reported by \cite{2003Martini} is (assuming Poisson distribution) $\sqrt{N_{\rm spi}}/N \simeq 4.6/28 \simeq 16\%$ and for our sample the same estimation gives uncertainty of $\sim16\%$. The difference may also result from the different number of sources or range of galaxy distances between these two, still relatively small, samples.}
 
 All of our tightly wound (tw) nuclear spirals belong to galaxies lacking a large-scale bar, in agreement with the results of \citet{2003Martini}. It is intriguing to note that one of our three grand-design nuclear spirals, in UGC~3789, also appears to lack a large-scale bar. While \cite{2003Martini} and other early works \citep{2000Englmaier, 2004Martini} argued that grand-design nuclear spirals are associated mostly with strongly barred galaxies, our findings are fully consistent with later works, e.g., \cite{2006Peeples} and \cite{2012Kim}, who find no evidence of such a correlation on the smallest scales. \cite{2012Kim} also predict that grand-design nuclear spirals should preferentially occur with circular ($x_2$-type) kpc-scale rings -- such a ring is indeed seen in UGC~3789. The galaxy also contains a pair of rings in an 8-shaped structure at scales of $\sim 20$~kpc. Their presence suggests that the grand-design nuclear behavior may have been caused by a recent merger. We note, however, that such ring structures are not necessarily a result of interactions and may also arise from the intrinsic dynamics of this galaxy \citep{2004Kormendy}. The fraction of tightly-wound spirals in our sample (5 out of 18 galaxies, i.e., $\sim28\%$) is consistent with that reported by \citet{2007SimoesLopes}, who find them in $\lesssim 25\%$ of their sample of early-type (elliptical and lenticular) galaxies. Whether nuclear dust morphology depends on galaxy type is, so far, uncertain \citep{2002Xilouris, 2003Martini, 2007SimoesLopes}.
 
 %have found for their sample of $75$ disk galaxies that only the large, kpc-scale spiral structure correlates with bar strength. They did not see evidence of any such correlation for their small nuclear spirals (which they required to extend all the way into the nucleus). This is consistent with the conclusions of \cite{2012Kim} who, based on hydrodynamical simulations, linked the dependence of nuclear spiral's pitch angle on bar strength to the speed of evolutionary unwinding of these spirals. This process occurs more slowly for weakly-barred galaxies and hence, tightly-wound nuclear spirals are more commonly observed there. No such effect is expected for the prevalence of nuclear grand-design spirals. \cite{2012Kim} also predict that grand-design nuclear spirals should preferentially occur with circular ($x_2$-type) kpc-scale rings, as eccentric rings would quickly destroy any grand-design spiral structure. Indeed, a round kpc-scale ring does accompany the grand-design spiral in UGC~3789. The galaxy also contains a pair of rings in an 8-shaped structure at scales of $\sim 20$~kpc. Their presence suggests that the grand-design nuclear behavior may have been caused by a recent merger. We note, however, that such ring structures are not necessarily a result of interactions and may also arise from the intrinsic dynamics of this galaxy \citep{2004Kormendy}.} 
 
 \section{Results: angular momenta}\label{sect:results_angmom}
 
 Table~\ref{tab:orient} gives the orientation of the angular momentum of the galactic masers, nuclear regions and kpc-scale disks; with the latter two based on the \texttt{ellipse} fits shown in Figures~\ref{fig:3rdPage1} and\ \ref{fig:3rdPage_ESO558}~--~\ref{fig:3rdPage_UGC6093}. Note that nuclear bulges are excluded from this analysis.
 
 \subsection{Relative orientation of angular momenta}
 \label{sect:dissRelOrient}
 
 The 3D angles and projected PA differences between the angular momenta of different sub-components in our analysis are given in Table~\ref{tab:angles}.
 
 \begin{table*}
 \centering
 \begin{threeparttable}
  \caption{Relative orientation of structures.}
 \renewcommand*{\arraystretch}{1.2}
 \begin{tabular}{|c|l|cc|cc|cc|}
 \hline
 & \multirow{2}{*}{Object} & \multicolumn{2}{c|}{Galaxy -- inner region} & \multicolumn{2}{c|}{Galaxy -- maser} & \multicolumn{2}{c|}{Inner region -- maser} \\
 & & $\Delta$PA [deg] & 3D angle [deg] & $\Delta$PA [deg] & 3D angle [deg] & $\Delta$PA [deg] & 3D angle [deg] \\
 (1) & \multicolumn{1}{c|}{(2)} & (3) & (4) & (5) & (6) & (7) & (8) \\
 \hline
 \multirow{13}{*}[2.5em]{\begin{sideways}This work\end{sideways}}
 & ESO558$-$G009 & 2, 178  & 15, 43, 137, 165 & 73 & 74 & 71, 109  &  73, 107 \\
 & J0437+2456 & --- & --- & 11 & 36 & --- & --- \\
 & Mrk1029 & 50, 130  & 40, 86, 94, 140 & 124 & 115 & 6, 174  &  33, 147 \\
 & Mrk1210 & 83, 97  & 30, 34, 146, 150 & 63 & 92 & 34, 146  &  58, 79, 101, 122 \\
 & NGC5495 & 62, 118  & 35, 58, 122, 145 & 29, 151 & 57, 123 & 33, 147  &  66, 114 \\
 & NGC5728 & 1, 179  & 45, 87, 93, 135 & 27 & 36 & 28, 152  &  72, 108 \\
 & NGC5765b & 90  & 49, 131 & 100 & 100 & 10, 170  &  58, 68, 112, 122 \\
 & UGC3193 & 1, 179  & 7, 43, 137, 173 & 20 & 27 & 19, 161  &  31, 149 \\
 & UGC6093 & --- & --- & 70 & 77 & --- & --- \\
\hline
\multirow{8}{*}[-0.8em]{\begin{sideways}\cite{2013Greene}\end{sideways}}
 & IC2560 & 3, 177  & 14, 70, 110, 166 & 89 & 89 & 88, 92  &  89, 91 \\
 & NGC1194 & 18, 162  & 22, 85, 95, 158 & 17 & 32 & 1, 179  &  46, 56, 124, 134 \\
 & NGC2273 & 45, 135  & 38, 88, 92, 142 & 113 & 101 & 68, 112  &  68, 75, 105, 112 \\
 & NGC2960 & 5, 175  & 16, 84, 96, 164 & 175 & 131 & 0, 180  &  33, 35, 145, 147 \\
 & NGC3393 & 14, 166  & 10, 88, 92, 170 & 164 & 134 & 2, 178  &  47, 133 \\
 & NGC4388 & 15  & 19 & 17 & 21 & 32  &  32 \\
 & NGC6264 & 13, 167  & 13, 86, 94, 167 & 105 & 102 & 62, 118  &  71, 109 \\
 & NGC6323 & --- & --- & 165 & 154 & --- & --- \\
 & UGC3789 & 25, 155  & 28, 84, 96, 152 & 68 & 72 & 43, 137  &  62, 63, 117, 118 \\

 \hline
\end{tabular}
\begin{tablenotes}\footnotesize
 \item \textbf{Notes:} Column 1 -- data sample. Column 2 -- host galaxy name. Columns 3-4 -- projected position angle difference ($\Delta$PA) and 3D angle between the angular momenta of the galaxy as a whole and the nuclear region. Columns 5-8 -- orientation of the maser disk relative to the galaxy as a whole (columns 5-6) and the nuclear region (7-8). In each case, all the angles allowed by the data are given (see discussion on degeneracy of relative orientation angles in Sect.~\ref{sect:orient}).
\end{tablenotes}
\label{tab:angles}
\end{threeparttable}
\end{table*}
 
 To quantify the trends visible in the relative orientation of the angular momenta in our samples we use one-sample Kolmogorov-Smirnov (KS) tests. This common statistical test allows us to assess whether a given sample is drawn from a specified distribution.
 
 In our implementation we use KS tests to assess whether the relative orientation of the angular momenta between two subcomponents (e.g., the maser disk and galaxy as a whole) could have been drawn from a random distribution between some limiting values. In the case of 3D angles $\psi$, the randomly distributed sample has a distribution $\frac{dN}{d\psi} \propto \sin(\psi)$. The Cumulative Distribution Function (CDF) of such a distribution with $\psi \in [\psi_1, \psi_2]$ is given by:
 \beq CDF_{\rm 3D}(\psi) = \begin{cases}
 0 & , \psi < \psi_1; \\
 \frac{\cos \psi_1 - \cos \psi}{\cos \psi_1 - \cos \psi_2} & , \psi_1 < \psi < \psi_2; \\
 1 & , \psi_2 < \psi. \\
 \end{cases} \label{eq:CDF_3D}\eeq
 For the projected PA differences between vectors, the random distribution is flat. The CDF of such a distribution with $\textrm{PA} \in [\textrm{PA}_1, \textrm{PA}_2]$ is simply:
 \beq CDF_{\Delta\textrm{PA}}(\textrm{PA}) = \begin{cases}
 0 & , \textrm{PA} < \textrm{PA}_1; \\
 \frac{\textrm{PA} - \textrm{PA}_1}{\textrm{PA}_2 - \textrm{PA}_1} & , \textrm{PA}_1 < \textrm{PA} < \textrm{PA}_2; \\
 1 & , \textrm{PA}_2 < \textrm{PA}. \\
 \end{cases} \label{eq:CDF_PA} \eeq
 The KS-statistics and p-values resulting from comparisons of these CDFs with CDFs of respective data samples are given in Sections~\ref{subsect:galincl}--\ref{subsect:regmas}, where applicable, and summarized in Table~\ref{tab:KS}.

 \begin{table}
  \centering
  \begin{threeparttable}
  \caption{Results of the KS tests.}
  \renewcommand*{\arraystretch}{1.1}
  \begin{tabular}{|cc|c|c|c|cc|}
    \hline
    \multicolumn{2}{|c|}{Species} & Angle & Limit & N & KS-stat & p-value \\
    (1) & (2) & (3) & (4) & (5) & (6) & (7) \\
    \hline
    Gal & Nuc & $\Delta$PA & $30\degree$ & 15 & $0.33$ & $0.05$ \\ 
Gal & Nuc & $\Delta$PA & $90\degree$ & 15 & $0.40$ & $0.01$ \\ 
Gal & Nuc(S) & $\Delta$PA & $30\degree$ & 7 & $0.43$ & $0.11$ \\ 
Gal & Nuc(S) & $\Delta$PA & $90\degree$ & 7 & $0.29$ & $0.50$ \\ 
Gal & Nuc(L) & $\Delta$PA & $30\degree$ & 8 & $0.33$ & $0.27$ \\ 
Gal & Nuc(L) & $\Delta$PA & $90\degree$ & 8 & $0.59$ & $3\times 10^{-3}$ \\ 
\hline
Nuc & Mas & $\Delta$PA & $90\degree$ & 15 & $0.29$ & $0.13$ \\ 
Nuc(S) & Mas & $\Delta$PA & $90\degree$ & 7 & $0.24$ & $0.83$ \\ 
Nuc(L) & Mas & $\Delta$PA & $90\degree$ & 8 & $0.44$ & $0.06$ \\ 
Nuc & Mas & 3D & $90\degree$ & 15 & $0.23$ & $0.37$ \\ 
Nuc(S) & Mas & 3D & $90\degree$ & 7 & $0.25$ & $0.70$ \\ 
Nuc(L) & Mas & 3D & $90\degree$ & 8 & $0.29$ & $0.42$ \\ 
\hline
Gal$^*$ & Mas & $\Delta$PA & $90\degree$ & 18 & $0.23$ & $0.24$ \\ 
Gal$^*$ & Mas & 3D & $180\degree$ & 17 & $0.20$ & $0.46$ \\ 
\hline

  \end{tabular}
\begin{tablenotes}\footnotesize
 \item \textbf{Notes:} The results of one-sample KS tests as described in Sect.~\ref{sect:dissRelOrient}. Columns~1-2: structures in the galaxy whose angular momenta orientations are being compared. Gal -- angular momentum of the galaxy as a whole, Gal$^*$ -- the same limited to galaxies with orientation fixed using rotation curves or the relative dust lane prominence method, Nuc -- the nuclear regions (in some cases divided into S -- those with $r<200$~pc and L -- with $r\ge200$~pc), Mas -- the megamaser disk. Column~3: type of relative angle: $\Delta$PA is the position angle difference (as defined in the text), 3D denotes the 3D angles between angular momenta. Column~4: upper limit of the range of the respective angle in which the control distribution is random (the lower limit in all cases is $0$). Column 5 -- number of sources in the sample. Columns~6-7 -- KS-statistic (column~6; maximal difference between CDFs, or Cumulative Distribution Functions, of the sample and test distribution) and p-values (column~7; likelihoods of the sample being drawn from the test distribution) resulting from each of the tests.
\end{tablenotes}
\label{tab:KS}
\end{threeparttable}
\end{table}
 
 \subsubsection{Masing disks are randomly oriented relative to kpc-scale disks}\label{subsect:galincl}
 
 \begin{figure}
  \centering
  \includegraphics{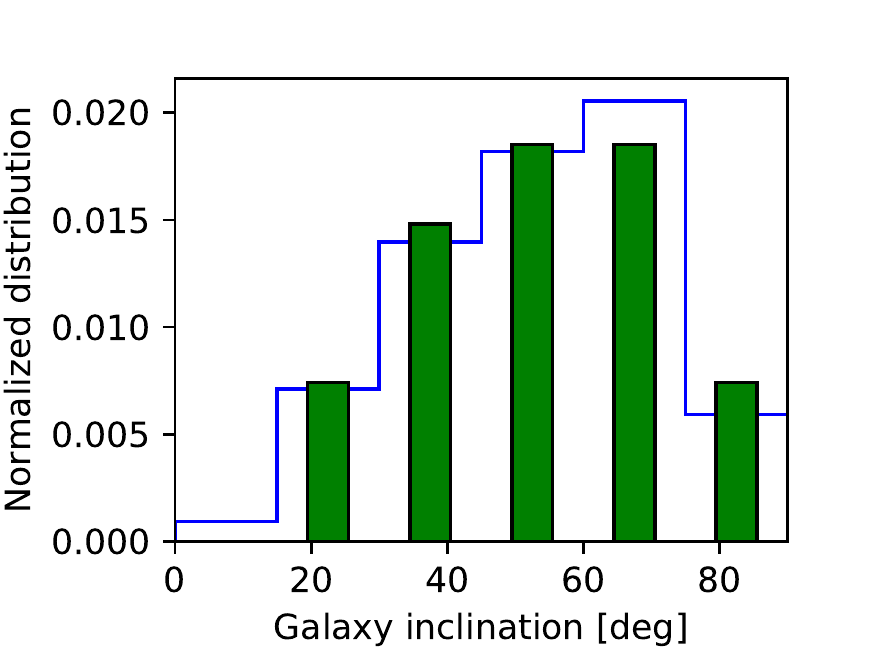}
 \caption{Comparison of galaxy inclinations in our sample with SDSS data, histograms are normalized so that the sum of all bins in each series multiplied by the width of the bins ($15\degree$ in both cases) is equal to 1. The inclination distribution of 788~disk galaxies from SDSS~DR4 \citep[][see Sect.~\ref{subsect:galincl} for details of galaxy choice from their sample]{2010Nair} is shown as the blue curve. For the sake of the comparison presented here, we ignore the information about spiral arms in our galaxies and present the lower of two possible inclination angles allowed in such case (see main text for discussion) as narrow green bars. In all other analyses in this work, the unambiguous galaxy inclination angle in $[0\degree,180\degree]$ range is used.}
 \label{fig:galincl}
\end{figure}
 
 Let us start with the largest scales in our analysis, the orientation of the galaxy on kpc scales (columns 7-9 in Table~\ref{tab:orient}). We first compare inclinations of our galaxies with those reported for disk galaxies by \cite{1997Braatz} and the Sloan Digital Sky Survey Data Release~4 (SDSS~DR4). For our sample of 18~galaxies the assumption of trailing spiral arms allows us to place each galaxy's inclination within the full $[0\degree,180\degree]$ range ($0\degree$ meaning that the angular momentum pointing towards Earth and $180\degree$ -- away from it). Since inclinations reported in \cite{1997Braatz} and SDSS~DR4 are based on ellipticity only, each of them allows two possible angular momentum orientations for the galaxy: for each $i_{\rm lit}$, $180\degree-i_{\rm lit}$ is also possible (see discussion in Sect.~\ref{sect:orient}). For the sake of the comparison presented here, we ignore the information about spiral arms winding directions in our galaxies and consider the lower of two possible inclination angles allowed in such case. In all other analyses in this work, the unambiguous galaxy inclination angle in the $[0\degree,180\degree]$ range is used.
 
 From comparison of our results shown in Fig.~\ref{fig:galincl} with Fig.~9 of \cite{1997Braatz} we can see that the maser galaxies in our sample appear to follow the inclination distribution of the entire population of galaxies -- both with and without detected megamasers. We check this result by performing a two-sample KS test with a sample of 788 disk galaxies from SDSS~DR4, using the galaxy morphology classification of \citet{2010Nair}. From their sample of 14~034 galaxies (limited to redshifts of $0.01<z<0.1$ and g'-band magnitudes brighter than 16~mag) we have chosen reliably classified (T-Type flag $=0$) galaxies of types Sa-Sd (T-Type of 1-7) and without signs of interaction (interaction flag ``$2^1$'' in their notation). The two-sample KS test of these 788 sources with our data results in a KS-statistic (maximal difference between CDFs of the two samples) of $0.12$ and p-value (likelihood of the two samples being drawn from the same distribution) of $0.94$. The inclinations of our galaxies are then strongly representative of the entire population of disk galaxies. Thus, the orientation of the maser disk is unlikely to depend on the orientation of the host galaxy.
 
 Such a non-correlation excludes the possibility that the gas inflow within the central pc is associated with non-axisymmetries in the large-scale structure of the galaxy. As this result is further supported by considerations below, we discuss it in more detail in Sect.~\ref{subsect:regmas} and Sect.~\ref{sect:discussion}.

 \begin{figure}
  \centering
  \includegraphics{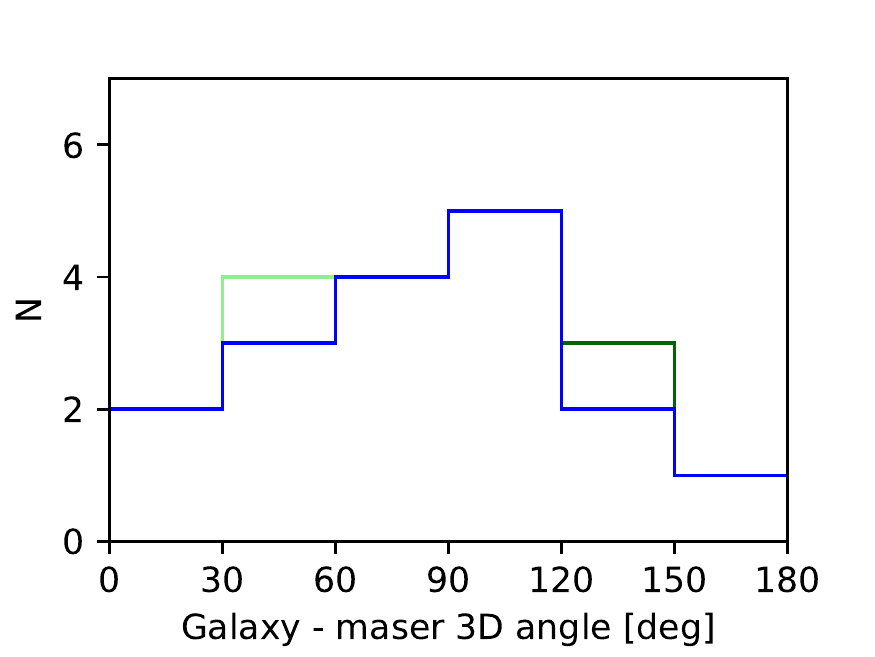}
  \caption{Galaxy-maser orientation comparison, 3D angle (angle between angular momenta of the two regions). Blue curve shows the histogram for sources with known rotation curves or orientation fixed using the relative dust lane prominence method. Light and dark green curves correspond to addition of the two possible 3D angles for NGC~5495. The distribution is consistent with random (see text, Sect.~\ref{subsect:galincl}).}% Bottom: position angle difference.}
  \label{fig:galmas}
 \end{figure}
 
 We also directly compare the kpc-scale structures with the orientation of the maser disks. Figure~\ref{fig:galmas} shows the histograms describing the relative orientation of the galaxy as a whole and the maser disk. In the case of the 3D angles for the galaxies with orientation fixed by rotation curves or relative prominence of dust lanes (blue curve), the distribution is consistent with being random: for the 3D angles between $\psi_1 = 0$ and $\psi_2=180\degree$, KS-statistic~$ \simeq 0.20$ and p-val~$ \simeq 0.46$.
 
 \subsubsection{Large nuclear regions align with kpc-scale galaxy while small-sized ones are oriented randomly}\label{subsec:galreg}
 
 \begin{figure}
  \centering
  \includegraphics{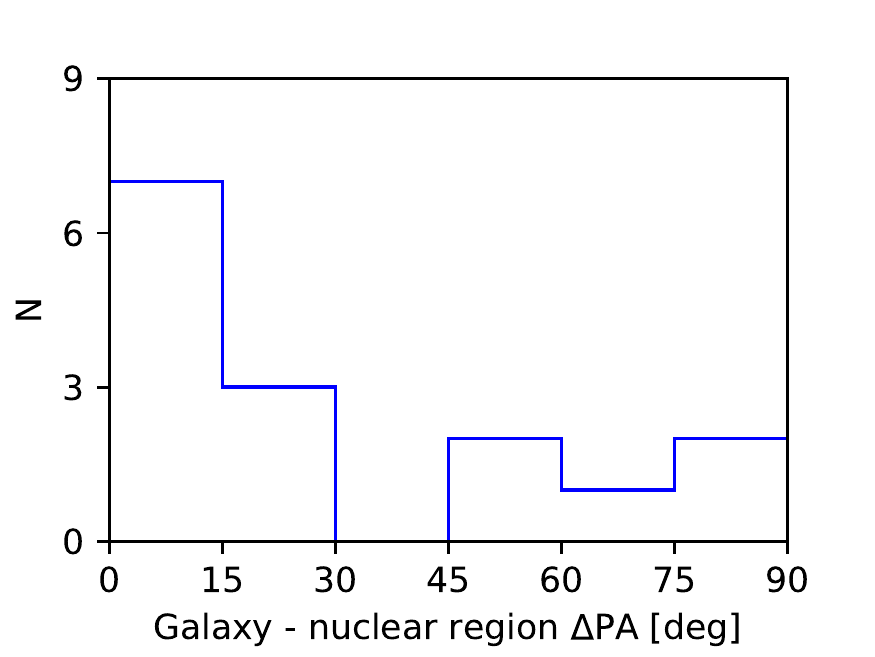}
  \includegraphics{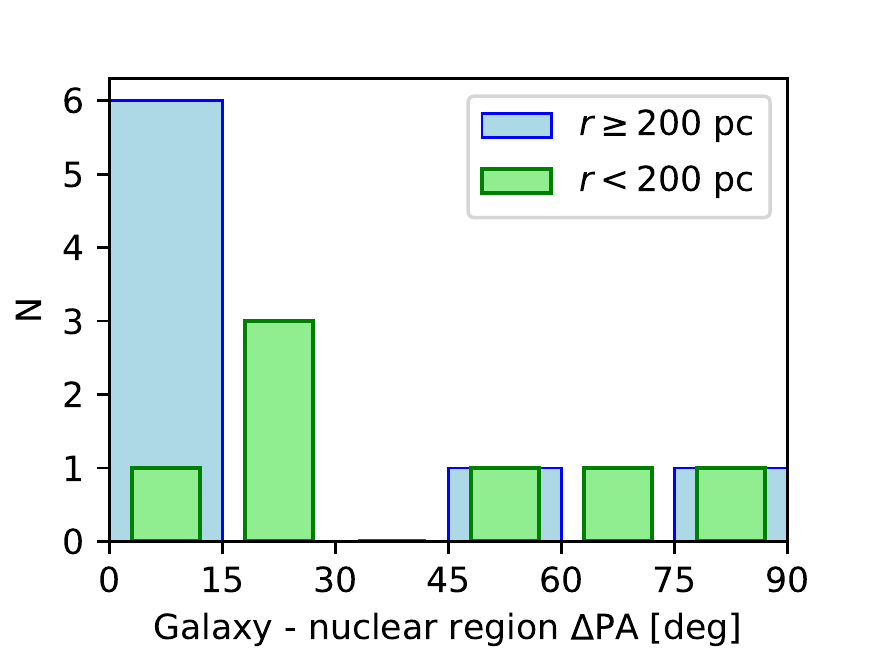}
  \caption{Position angle difference between the angular momenta of the galaxy and its nuclear region. Only the smaller of two possible angles is shown (the other one is the result of subtraction of the shown angle from $180\degree$). Top: full sample. Bottom: small ($r<200$~pc) and large ($r\ge200$~pc) nuclear regions shown separately as (narrow) green and (full-width) blue bars. The observed misalignment arises due to the smaller nuclear regions (see text, Sect.~\ref{subsec:galreg}).}
  \label{fig:galregPA}
 \end{figure}

 %While most nuclear regions seem to be aligned in PA with their host galaxies (consistently with the results of, e.g.,\citealt{2002Schmitt}), the KS test disfavours the hypothesis that all nuclear regions tend to do so (out of $1000$ galaxies $683\pm176$ are expected to have PA difference lower than $30\degree$; for KS-statistic and p-value see Table~\ref{tab:KS}).

 We next compare the orientation of the kpc-scale galaxy to the 100~pc scale nuclear regions. Overall, the nuclear regions have too broad a distribution of projected $\Delta$PA to be aligned -- the p-value for comparison with a fully random distribution is only $0.01$ (see Fig.~\ref{fig:galregPA} and Table~\ref{tab:KS}). There seems to be a weak hint of misalignment growing with decreasing scale of the structure. If we cut our random control distribution at a maximum angle of $\Delta$PA$_2 = 30\degree$, we see that the large nuclear regions are consistent with this aligned sample, while for smaller nuclear regions a fully random distribution is much more likely (see Table~\ref{tab:KS}). While this division is arbitrary, the difference suggests that the nuclear structure progressively misaligns from the large-scale disk. We caution, however, that the evidence for this behavior is marginal.
 
 %This misalignment arises due to small ($r<200$~pc) nuclear regions -- the best fit of our simplified model in their case corresponds to the projected position angle difference being completely random (with KS-statistic and p-value of $0.29$ and $0.50$). \textbf{To see if the large nuclear regions tend to align with their corresponding kpc-scale disks, we compare the position angle differences between their angular momenta and those of their kpc-scale disks with a random distribution with $\Delta$PA$<30\degree$. For large nuclear regions, the p-value for $\Delta$PA$<30\degree$ is $0.36$; while for $\Delta$PA$<90\degree$, i.e., random orientation, the p-value is $<0.01$. Even though the $30\degree$ boundary is arbitrary, we clearly see preference for alignment with kpc-scale galaxy for large nuclear regions. Note that alignment is also not ruled out for the small nuclear regions, as the p-value for $\Delta$PA$<30\degree$ is in their case $0.11$. However, as the full sample is inconsistent with having $\Delta$PA$<30\degree$, we expect the two distributions to be different. Since it is unlikely that the small nuclear regions align with kpc-scale galaxy better than the large ones, we conclude that their true distribution, if not fully random, must be at least broader than that of the large nuclear regions.}
 
 While the boundary between smaller and larger nuclear regions of $200$~pc, on which the numerical results presented above are based, is somewhat arbitrary (see Sect.~\ref{sect:classification}), there is a clear change towards more misalignment as the structure size approaches $<100$~pc scales.
 
 We then observe a build-up of misalignment with regard to the large-scale galactic disk as we go deeper into the galactic nucleus. If this misalignment continues to grow, once we reach $\sim0.1$~pc scales, at which megamaser disks reside, the gaseous disk is completely randomly oriented with regards to the galactic disk (see next subsection).
 
 \subsubsection{Nuclear regions are randomly oriented with regard to the megamaser disks}\label{subsect:regmas}
 
 \begin{figure}
  \centering
  \includegraphics{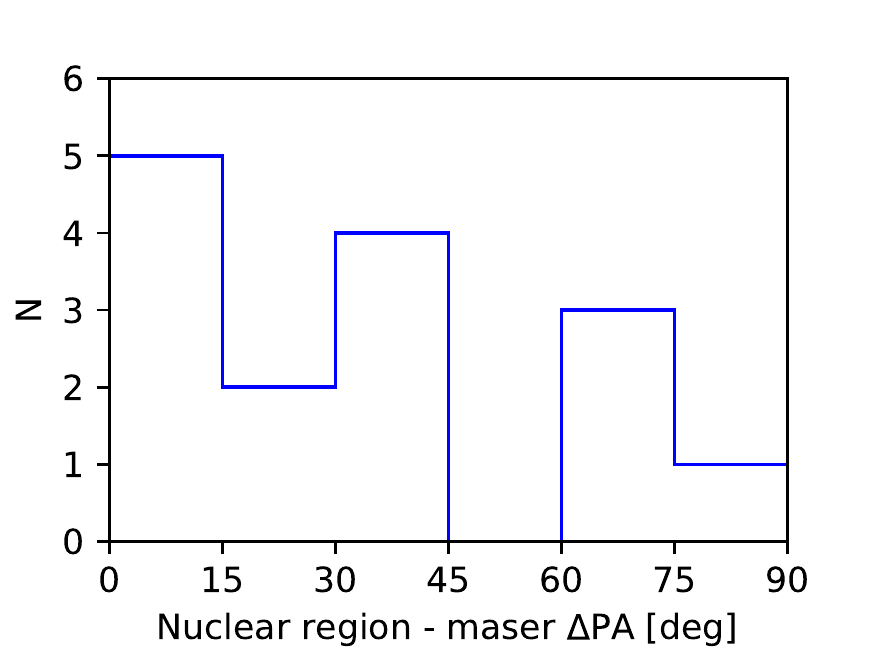}
  \includegraphics{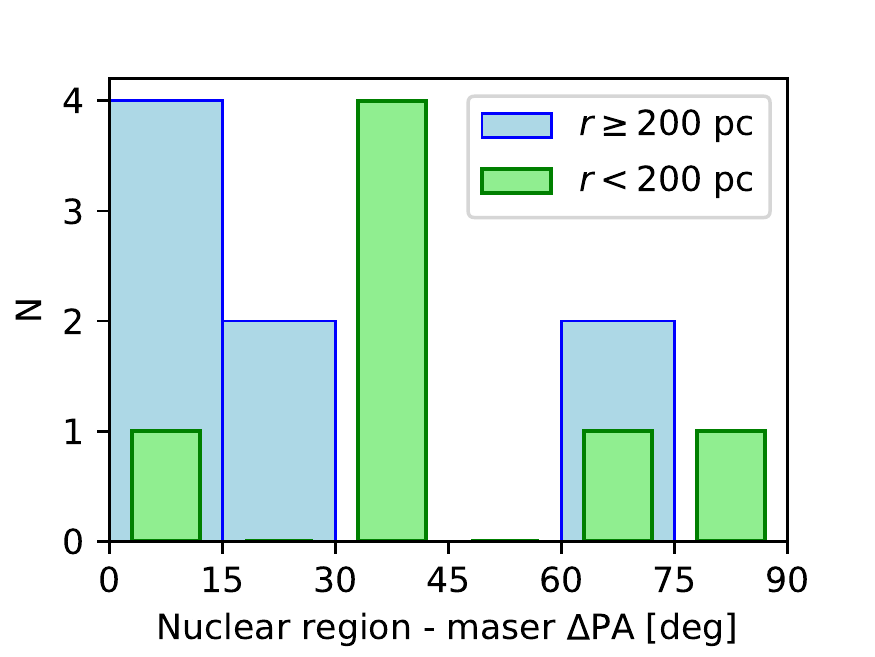}
  \caption{Position angle difference between nuclear region and maser angular momentum orientation. Only the smaller of two possible angles is shown (the other one is the result of subtraction of the shown angle from $180\degree$). Top: full sample. Bottom: small ($r<200$~pc) and large ($r\ge200$~pc) nuclear regions shown separately as (narrow) green and (full-width) blue bars. Both small and large nuclear regions are consistent with being randomly oriented with regard to their megamaser disks (see text, Sect.~\ref{subsect:regmas}).}
  \label{fig:regmasPA}
 \end{figure}

 \begin{figure*}[h!]
 \centering
 \begin{tabular}{cc}
  \includegraphics{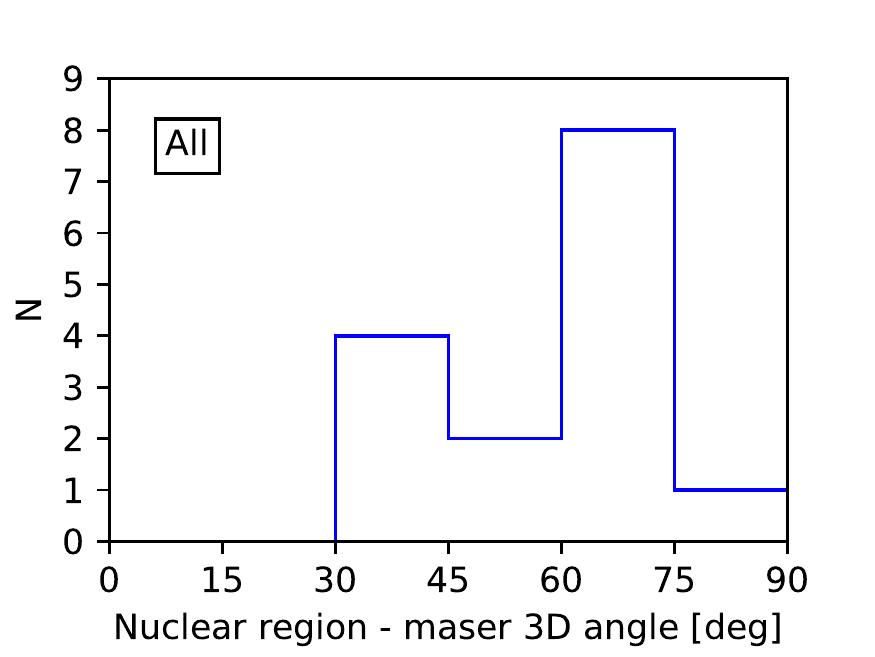} & \includegraphics{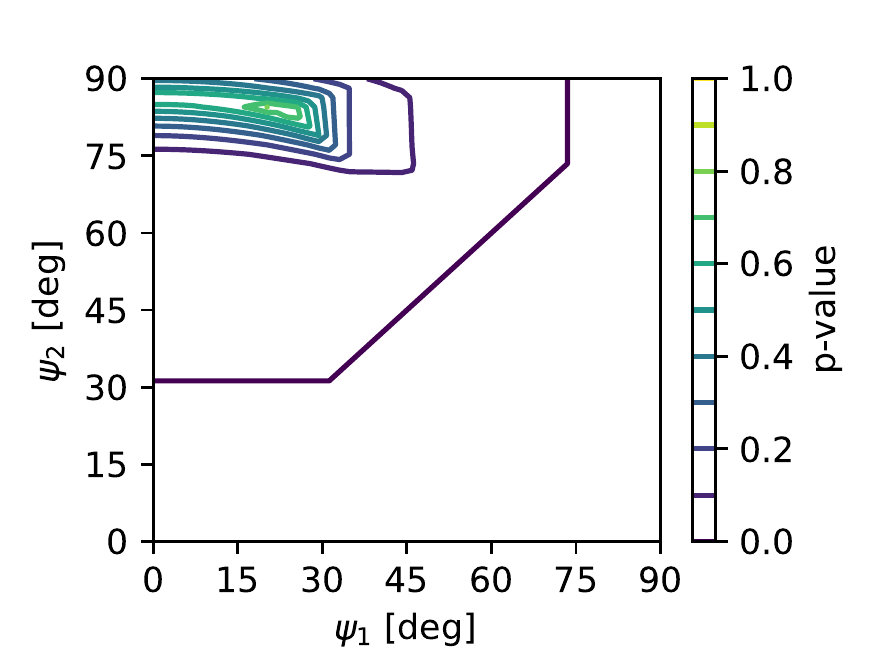} \\
  \includegraphics{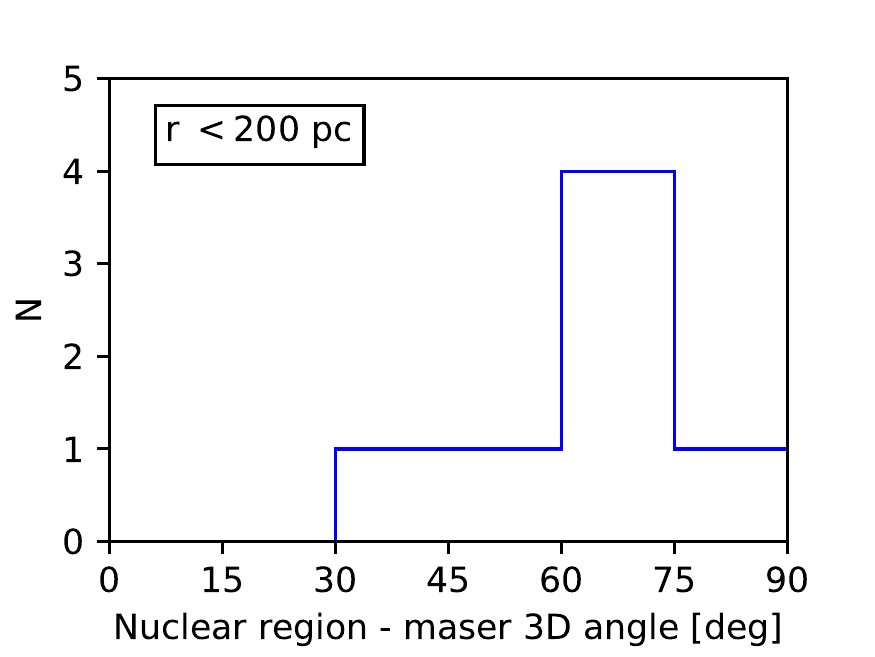} & \includegraphics{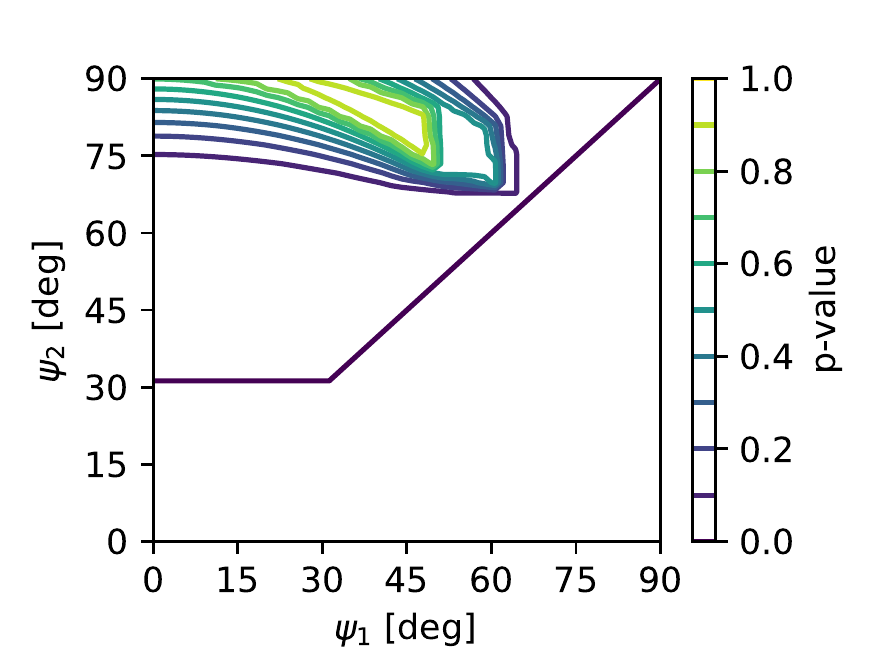} \\
  \includegraphics{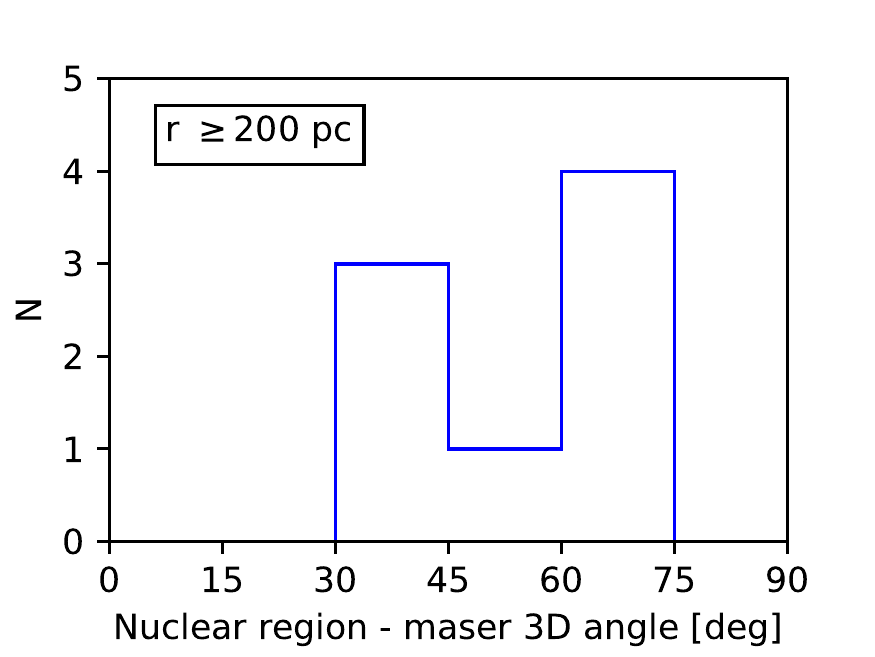} & \includegraphics{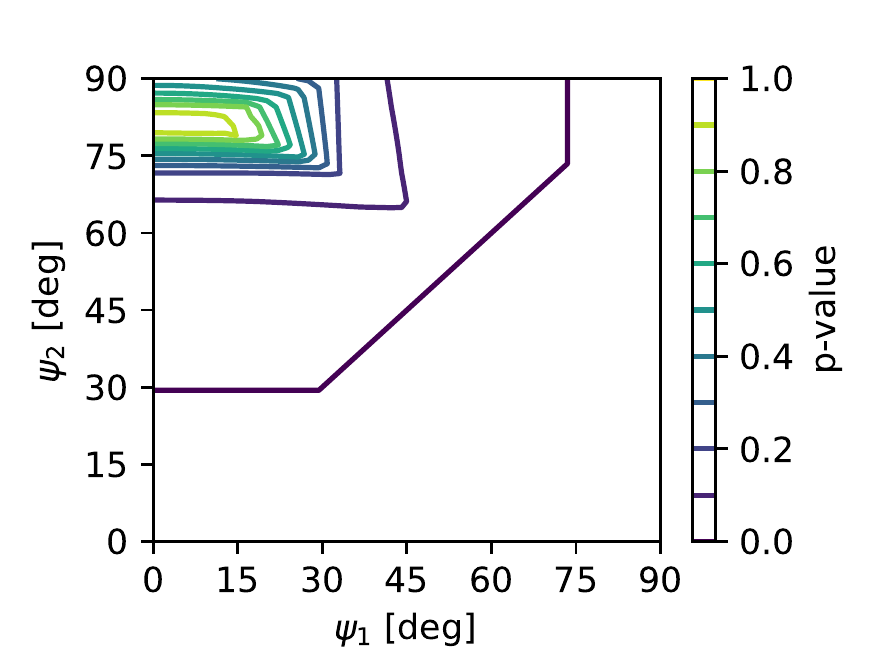}
 \end{tabular}
  \caption{Left: histograms of 3D angles between angular momentum orientation of the nuclear region and the masing disk for all sources (top), nuclear regions smaller than $200$~pc (middle) and those larger than $200$~pc (bottom). Only the lower halve of the full distribution ($\psi \in [0,180\degree]$) is shown, limiting the data to the lower of two possible 3D angles (see main text for discussion). Right: contour maps of p-values obtained by comparison of respective samples with random distribution of 3D angles between $\psi_1$ and $\psi_2$ by means of one-sample KS tests. The regions of highest p-value (light green) in the right column correspond the regions in parameter space favored by the data. The distribution is consistent with random ($\psi_1 = 0$, $\psi_2 = 90\degree$) for both small and large nuclear regions (see text, Sect.~\ref{subsect:regmas}).}
  \label{fig:regmas3D}
 \end{figure*}

 Now we compare the 100~pc scale nuclear regions to objects on the smallest scales in our analysis - the maser disks $\sim 0.5$~pc from the SMBH. Figure~\ref{fig:regmasPA} shows the difference in projected position angle ($\Delta$PA) between the angular momenta of the nuclear region and the maser disk. For the entire sample, the $\Delta$PA distribution is consistent with being random, and the K-statistic and p-value give $(0.29,0.13)$. Then, the left column of Fig.~\ref{fig:regmas3D} shows the distribution of 3D angles between the angular momenta of the nuclear region and the maser disk. Note that due to the maser disk being edge-on, only two such angles result from the data\footnote{In the cases where the inclination of the maser disk is known to be different from $90\degree$, the resulting 3D angle possibilities consist of two pairs of close angles (see Table~\ref{tab:angles}). In these cases, averages of those pairs are used here as the two possible 3D angles.}. The two possible angles are symmetric with respect to $90\degree$, i.e., for each 3D angle $\psi$, $180\degree-\psi$ is also possible\footnote{For the purpose of this part of our analysis, we ignore the fact that for NGC~4388 the orientation of the inner nuclear region is fixed by \cite{2014Greene} and instead assume that also there two 3D angles are possible: $58\degree$ (the true result) and $180\degree-58\degree$. This allows us to analyze the entire sample in a homogeneous manner.} Since the two possible angles are symmetric with respect to $90\degree$, we only take into account the smaller of them and compare the resulting distribution with a random distribution of 3D~angles between $\psi_1$ and $\psi_2$. This is equivalent to assuming that the true distribution is also symmetric with respect to $90\degree$ and comparing only its lower half to a random distribution. Alternatively, we also tried to use both angles as two separate measurements (resulting in a measured distribution symmetric with respect to $90\degree$), but the results were then highly dependent on our assumption of symmetry with respect to $90\degree$ and, thus, less informative of the observed distribution.
 
 Contour plots of p-values resulting from comparison of a random distribution between 3D angles $\psi_1$ and $\psi_2$ with the distribution of the smaller of the possible 3D angles between nuclear regions and their respective masers are shown in Fig.~\ref{fig:regmas3D}. The distributions for the entire sample, as well as small and large nuclear regions separately, are consistent with random orientations: the KS-statistic and p-value pairs for $\psi_1 = 0$ and $\psi_2 = 90\degree$ are $(0.23, 0.37)$ for the entire sample, $(0.25, 0.70)$ for the small and $(0.29, 0.42)$ for the large nuclear regions. While the lower limit seems to be preferred at $\psi_1 > 0$, we attribute this to our distributions being biased away from $\psi_1 = 0$ (see Sect.~\ref{sect:orient}). Alternatively, an asymmetric distribution of true 3D angles $\{\psi_i\}$ would cause our distribution $\{\min(\psi_i, 180\degree-\psi_i)\}$ to be inconsistent with random, even if all nuclear regions do orient randomly with respect to their megamasing disks up to an angle of $\psi_2 \gtrsim 90\degree$.

 Our data fully support the hypothesis that the orientation of nuclear regions in our sample is not correlated with the orientation of the maser disks, regardless of the size of the nuclear region. This agrees well with previous results concerning the relative orientation of these structures that were obtained using jet orientation \citep{1984Ulvestad, 1999Nagar, 2000Kinney, 2002Schmitt, 2006Gallimore}, reflection spectroscopy \citep{2016Middleton}, narrow emission-line regions \citep{2013Fischer} and previous results drawn from H$_2$O megamaser orientations \citep{2013Greene}. In the framework of ``non-axisymmetric features all the way down'' model, this would mean that the non-axisymmetries responsible for the final infall of gas to the central supermassive black holes (expected to be aligned with the central accretion disk) are still unresolved in our observations.
 
 This result augments and adds to the findings of \cite{2003Martini}. Using \emph{HST} data, they found that the type of nuclear spiral at $\sim100$~pc does not correlate with AGN activity \cite[see also][]{2007SimoesLopes}. They conclude that the $100$~pc scales do not determine the accretion state of the galactic nucleus. This finding allowed them to constrain the duty cycle of AGN activity, requiring the duration of the active phases to be shorter than a few~Myr (the typical dynamical time at $100$~pc). Additionally, inflows have been observed to be associated with non-axisymmetries at these scales in observations using integral field spectroscopy \citep[IFS;][and references therein]{2009Davies, 2013Riffel, 2015Diniz}.
 
 The emerging picture is that while gas inflow feeding the SMBH passes through the $\sim100$~pc scales, the on/off state of AGN accretion is regulated closer to the galactic center. With the above in mind, some signs of galactic nuclear activity are visible in the dynamical state of gas at $100$~pc scales. There are hints that active galaxies have more centrally concentrated and more rotationally dominated central regions ($<200$~pc; \citealt{2013Hicks}), as well as a possibly higher molecular gas content within similar scales \citep{2016Izumi}.
 
 The nature of the on/off switch of galactic activity may be relevant to AGN feedback considerations. Currently, the numerical experiments regarding this process rarely resolve structures below $100$~pc and usually use more finely-resolved simulations at smaller scales (such as those of \citealt{2012Hopkins}) to generate prescriptions regulating the nuclear activity. If the true dynamics of SMBH feeding deviates from these prescriptions, the large-scale AGN feedback simulations may overestimate the overall accretion rate, and thus the total feedback. Precise timing of AGN on/off states may also be relevant to AGN feedback. However, this latter dependence may be non-trivial and its details are beyond the scope of this work.
 
\newpage 
 
 \section{Discussion and summary}\label{sect:discussion}
 
 We investigate $9$ new megamaser host galaxies, doubling the sample of \cite{2013Greene}. We use \emph{HST}/WFC3 data to identify the innermost resolved structures in each of them. In order to select those regions and extract their orientations, we use the ellipse-fitting algorithm of \cite{1987Jedrzejewski} and structure maps of \cite{2002Pogge}. We compare the orientation of each $100$--$500$~pc~scale nuclear region with the orientation of its host galaxy and the megamaser disk. We conclude that:
 \begin{itemize}
  \item The orientation of the galaxy relative to the masing disk is random -- confirming the separation of large-scale structure from the central gas infall onto the SMBH.
  \item The nuclear regions likely become increasingly misaligned from the large-scale galactic disk as their scale decreases. The ones smaller than $200$~pc are likely completely misaligned from the kpc-scale galaxy. There is a hint that larger nuclear regions are more aligned, as they show a preference for a maximum (arbitrarily chosen) 3D~angle of $30\degree$ over that of a fully random distribution.
  \item The orientation of the nuclear region relative to the masing disk is consistent with random, regardless of the size of the region -- we conclude that $\sim100$~pc~scale structures still do not directly couple to the inner accretion flow responsible for AGN activity, in agreement with previous studies.
 \end{itemize}
 
 We find that the structures within a galaxy become more and more misaligned as we approach the central supermassive black hole. While nuclear regions larger than $200$~pc in radius appear to align with their kpc-scale disk, the smaller ones become completely randomly oriented with respect to the large-scale structure. The central $\sim 0.1$~pc scale maser disk appears randomly oriented relative to both the nuclear regions and the kpc-scale galaxy. We therefore confirm the results of previous studies that both large-scale structures \citep{2000Kinney, 2006Gallimore, 2016Middleton} and those at $\sim100$~pc scales \citep{2002Schmitt, 2003Martini} do not couple directly to the central accretion flow. We conclude that the mechanism driving the final gas inflow onto the SMBH must operate closer to the black hole itself. Our findings are in agreement with the ``non-axisymmetric features all the way down'' model, where the nuclear structures are expected to increasingly misalign from the large-scale disks of galaxies as they transport gas further and further in \citep{2012Hopkins}. In the final parsec, where the gas reaches the megamaser disk and the AGN accretion disk, the flow may become completely misaligned from even the intermediate-scale structures.
 
 We favor a picture in which accretion is contingent on torques leading to shifts in angular momentum that randomize the orientations that we observe, but there are alternative explanations. One is that the gas on small scales has an external origin \citep[e.g.,][]{1984Bertola, 2006Morganti}. However, as mergers do not appear to be common enough to explain all nuclear activity \citep{2011Cisternas, 2012Kocevski}, this alternative cannot apply to all Seyfert galaxies. Very likely inflows mediated by both gravitational instability and galaxy interactions occur some of the time, as argued by \citet{2006Sarzi} for the case of lenticular and elliptical galaxies. In the case of the megamaser disk galaxies considered here, NGC~2960 is a likely merger remnant \citep{2016Laesker} and NGC~5765b is a member of an interacting pair of galaxies \citep{2016Gao}. UGC~3789 contains a pair of rings in an 8-shaped structure, but they are not necessarily a result of interactions and may arise from the intrinsic dynamics of this galaxy \citep{2004Kormendy}. Another possible channel of gas inflow comes from numerous interactions with small galaxies. However, as no galaxies other than NGC~2960 and NGC~5765b in our sample show obvious signs of recent disturbance \citep{2016Laesker, 2016Gao} this does not appear to be at work in the galaxies considered here.
 
 Warping of the nuclear galactic disk due to radiation pressure supplied by the AGN \citep{1996Maloney, 1997Pringle, 2000Gammie} is also worth consideration. At megamaser scales these processes result in warps observed in radio imaging of the maser spots (e.g., \citealt{1995Neufeld, 2003Greenhill, 2005Herrnstein}). This sub-parsec-scale deformations could misalign the observed masing disk direction with regard to the nuclear disk at $100$~pc scales, where the gas is unaffected by warping due to AGN radiation pressure.
 
 %When it comes to the type of structure at the nuclear regions, we observe a transition between a mixture of types (disks, rings, bulges, chaotic structures) at $150$-$300$~pc to almost exclusively flat environments at smaller scales. \textcolor{blue}{Not sure what you mean here?}
 
 We identify nuclear spirals in 8 out of 18 galaxies in our sample. One of them (UGC~3789) appears to be grand-design and resides in a galaxy without a large-scale bar, which has not been expected based on previous studies. We confirm that the tightly-wound nuclear spirals preferentially reside in non-barred galaxies.
 
 %We observe significant dust structure associated with features predicted by theoretical models following the 'stuff within stuff' paradigm. Therefore, we confirm this model to be a plausible explanation for the gas transport down to the inner $\sim 100$~pc of the galaxy. However, we conclude that these features are disconnected with the inner, pc-scale, accretion flow traced by the maser disk. Efforts are needed to probe structures an order of magnitude smaller in order to find regions that couple directly to the accretion disk; ALMA at full capacity will have the capability to accomplish this task for most of the galaxies in our sample.
 
 While megamaser host galaxies are an excellent target for an analysis of SMBH feeding mechanisms, the sample of such sources with optical imaging of sufficient angular resolution is very limited. As of now, only $\sim 20$ megamaser hosts have been imaged with \emph{HST} (\citealt{2010Greene}, this work). There are $34$ H$_2$O megamaser disks currently known \citep{2015Pesce, 2016vandenBosch}, so this number may double in the near future as additional observations are made. Moreover, upcoming instruments like the \emph{Atacama Large Millimeter/submillimeter Array} (\emph{ALMA}) and the \emph{James Webb Space Telescope} (\emph{JWST}) will allow to image the thermal gas at the galactic centers of megamaser host galaxies with unprecedented resolution and/or spectral capabilities. Using these data, which would directly trace the gas, our analysis could be repeated on a more statistically significant sample of sources. For now, however, our work presents the state of the art of optical imaging of megamaser host galactic nuclei -- we hope that not only will it broaden the understanding of gas inflows in galactic nuclei, but it will also be a useful starting point for similar endeavors in the future.
 
 %\newpage
 
 \section{Acknowledgements}
 J.E.G. acknowledges funding from NSF grant AST-1310405.
 
 We also acknowledge the incredible support of the late Fred Lo, whose undying enthusiasm and exacting scholarship drove all of us to do better research.
 
 The authors would like to thank the anonymous reviewer for helpful comments and suggestions which significantly improved this manuscript.
 
 This research has made use of the NASA/IPAC Extragalactic Database (NED) which is operated by the Jet Propulsion Laboratory, California Institute of Technology, under contract with the National Aeronautics and Space Administration (NASA).
 
 This research has made use of the Sloan Digital Sky Survey (SDSS) data. Funding for the SDSS has been provided by the Alfred P. Sloan Foundation, the Participating Institutions, the National Aeronautics and Space Administration, the National Science Foundation, the U.S. Department of Energy, the Japanese Monbukagakusho, and the Max Planck Society. The SDSS Web site is \url{http://www.sdss.org/}. The SDSS is managed by the Astrophysical Research Consortium (ARC) for the Participating Institutions. The Participating Institutions are The University of Chicago, Fermilab, the Institute for Advanced Study, the Japan Participation Group, The Johns Hopkins University, the Korean Scientist Group, Los Alamos National Laboratory, the Max-Planck-Institute for Astronomy (MPIA), the Max-Planck-Institute for Astrophysics (MPA), New Mexico State University, University of Pittsburgh, University of Portsmouth, Princeton University, the United States Naval Observatory, and the University of Washington.
 
 The authors greatly appreciate the availability of NASA's Astrophysics Data System Bibliographic Services (ADS), which have been extremely useful in preparation of this manuscript.
 
 \clearpage

% update the bibliography
 \bibliographystyle{apj}
 \bibliography{references}
 
 \newpage
 
 \appendix
 
 \section{Images, structure maps and \texttt{ellipse} profiles of the galaxies}\label{sect:3rdpageplots}
 
 \setcounter{figure}{0}
 \renewcommand{\thefigure}{A\arabic{figure}}
 
 \begin{figure}[H]
  \centering
  \makebox[\textwidth]{\includegraphics[scale=1.]{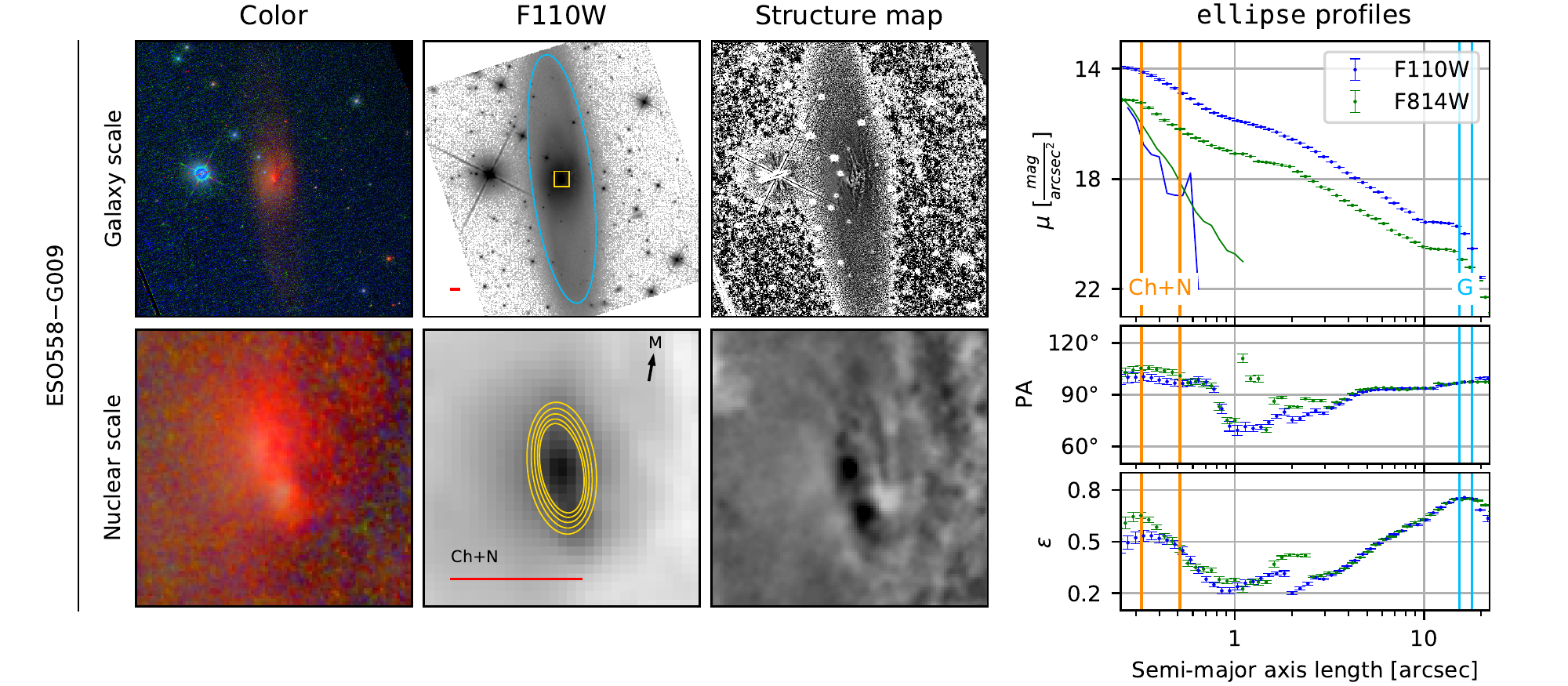}}
  \caption{Images and \texttt{ellipse} fits of ESO~558$-$G009. The order of images and plots, as well as the meaning of symbols and designations are the same as in Fig.~\ref{fig:3rdPage1}. In all images North is up and East is left. Column 1 (from left to right): false-color images (blue -- F335W, green -- F438W, and red -- F814W). Column 2: F110W image (color -- logarithmic scale for count rate). Column 3: structure maps constructed from the F814W image (logarithmic scale). In the F110W image in the top row (second column), ellipses following the galaxy-wide orientation are marked in blue and a yellow rectangle shows the region presented in the bottom row. In the F110W image in the bottom row, ellipses tracing the nuclear region are marked in yellow. The position angle of the maser disk angular momentum vector, perpendicular to the line of nodes of the masing disk, and the PA of the jet (if known) are shown in the upper right corner of this image as black arrows marked with ``M'' and ``J'', respectively. Jet orientation references for all 18 sources (where presented): \citet{1988Schommer, 1998Falcke, 2001Schmitt, 2009Mundell, 2010Xanthopoulos, 2012Yamauchi, 2013Sun} and the FIRST survey, \citealt{FIRST}. The red bar at the bottom of each F110W image is $1$~kpc in projected distance. Nuclear class (see Sect.~\ref{sect:classification}) is noted above the red bar on the nuclear-scale image. Column 4: Surface brightness ($\mu$, blue and green circles with horizontal bars indicating the angular range), position angle (PA) and ellipticity ($\epsilon$) profiles from \texttt{ellipse} for the F110W (blue) and F814W (green) images. Note that eccentricity $e=\sqrt{1-(1-\epsilon)^2}$. Vertical blue and orange lines limit the ranges of ellipse major axes used to extract the orientation of a galaxy as a whole and the nuclear region, respectively. The type of each structure is indicated. Blue and green solid lines on the surface brightness plot show \texttt{ellipse} fits to point sources in F110W and F814W images, respectively, approximating the point-spread function (PSF). The PSF profiles have been artificially scaled in brightness to optimize their visibility on the plots.}
  \label{fig:3rdPage_ESO558}
 \end{figure}
 
 \subsection{ESO~558$-$G009}

 ESO~558$-$G009 is a highly inclined Sb galaxy \citep{1982Lauberts} $115$~Mpc from Earth (luminosity distance, \citealt{2000Mould}). Two counterclockwise-wound spiral arms are visible in the F110W image. The central regions of ESO~558$-$G009 are obscured by a pronounced dust lane. We identify a nuclear region with radius $\sim0.4$~arcsec ($200$~pc). Its main identifying features in the \texttt{ellipse} profiles are a plateau in eccentricity in F110W between $0.3$ and $0.5$~arcsec, a break in surface brightness at its inner edge ($0.3$~arcsec) and a maximum in F814W PA profile at $0.4$~arcsec. The structure map reveals a dust structure associated with these features in \texttt{ellipse} fits, which supports the distinct character of this region. However, no clear morphological type can be assigned in this case.
 
 \newpage
 
 \begin{figure}
  \centering
  \makebox[\textwidth]{\includegraphics[scale=1.]{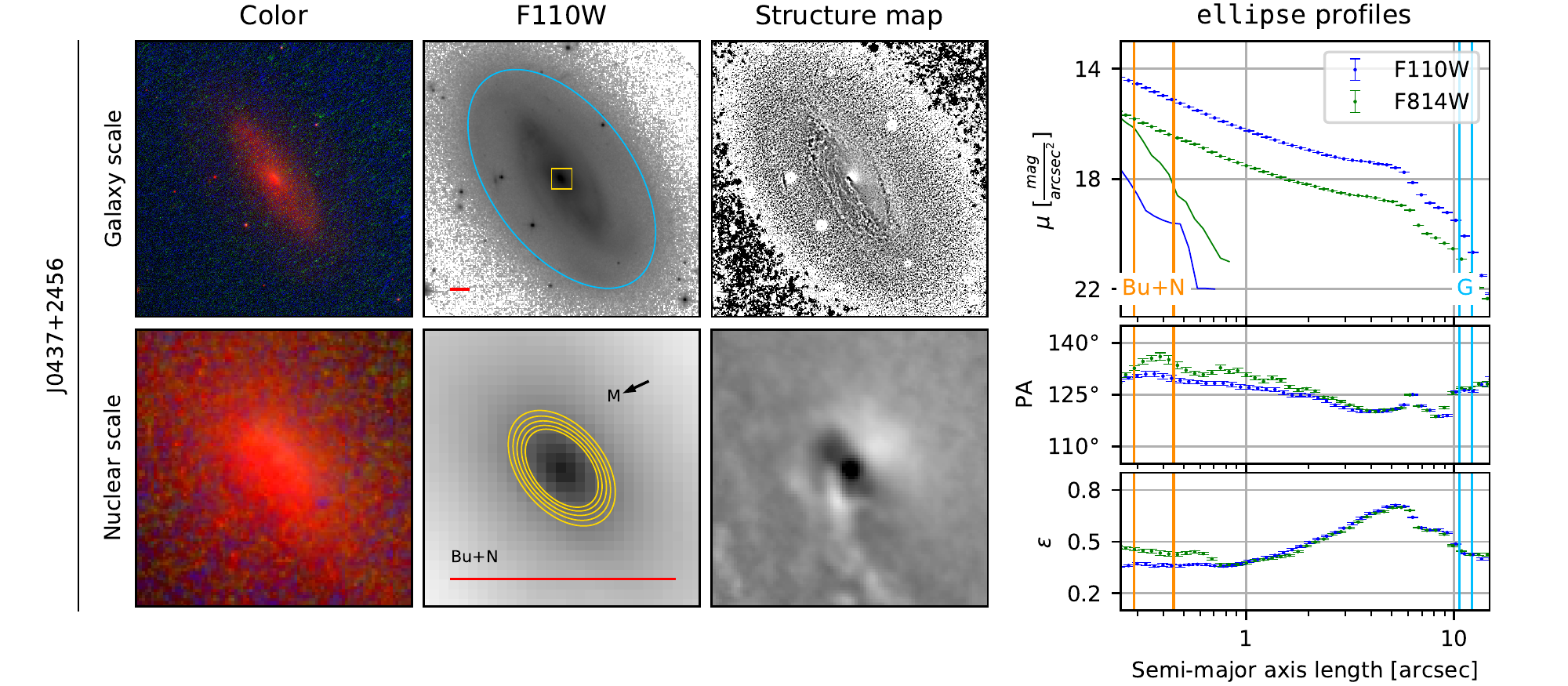}}
  \caption{Images and \texttt{ellipse} fits of J~0437+2456. The order of images and plots, as well as the meaning of symbols and designations are the same as in Fig.~\ref{fig:3rdPage1}.}
  \label{fig:3rdPage_J0437}
 \end{figure}
 
 \subsection{J~0437+2456}
 
  The luminosity distance to J~0437+2456 is $\sim 70$~Mpc \citep{2015Pesce}. It is a faint, probably Sb-type galaxy with no Hubble Type in the literature. The inclination is quite high, $\sim 60\degree$. Two counterclockwise-wound spiral arms are clearly visible extending from a ``boxy'' (peanut-shaped) bulge on larger scales. The central region of J~0437+2456 is marked by a flattened maximum in position angle profile at $\sim 0.4$~arcsec ($110$~pc), corresponding to a flat region in eccentricity. This region corresponds to an elliptical structure in the structure map and seems consistent with a small-scale bulge.
 
 %\begin{figure}
 % \centering
 % \makebox[\textwidth]{\includegraphics[scale=1.]{paper_Mrk1029_3rdPage.pdf}}
 % \caption{Images and \texttt{ellipse} fits of Mrk~1029. The order of images and plots, as well as the meaning of symbols and designations are the same as in Fig.~\ref{fig:3rdPage1}.}
 % \label{fig:3rdPage_Mrk1029}
 %\end{figure}
  
  \subsection{Mrk~1029}
  
  Mrk~1029 (Fig.~\ref{fig:3rdPage1}) is an Irr/S galaxy (first position data: \citealt{1981Kojoian}; morphological identification as spiral: 2MASS, \citealt{2MASS}) located $124$~Mpc from Earth (luminosity distance, \citealt{2000Mould}). It has a counterclockwise wound spiral structure (arms winding counterclockwise from inside out) consisting of two faint kpc-scale spiral arms. The galaxy seems to be moderately inclined. As the nuclear region in the galaxy, we identify a $\sim 0.4$~arcsec ($220$~pc) structure, where the position angle profile flattens and the ellipticity forms a distinct region with an inward increase. A feature in the surface brightness profile can also be identified in both the F814W and F110W filters. Spatially, the selected nuclear region follows a dust structure clearly visible on the structure map. An interesting feature of this galaxy is the absence of any structure in the dust on scales larger than $\sim 0.5$~kpc.
  
  \newpage
 
 \begin{figure}
  \centering
  \makebox[\textwidth]{\includegraphics[scale=1.]{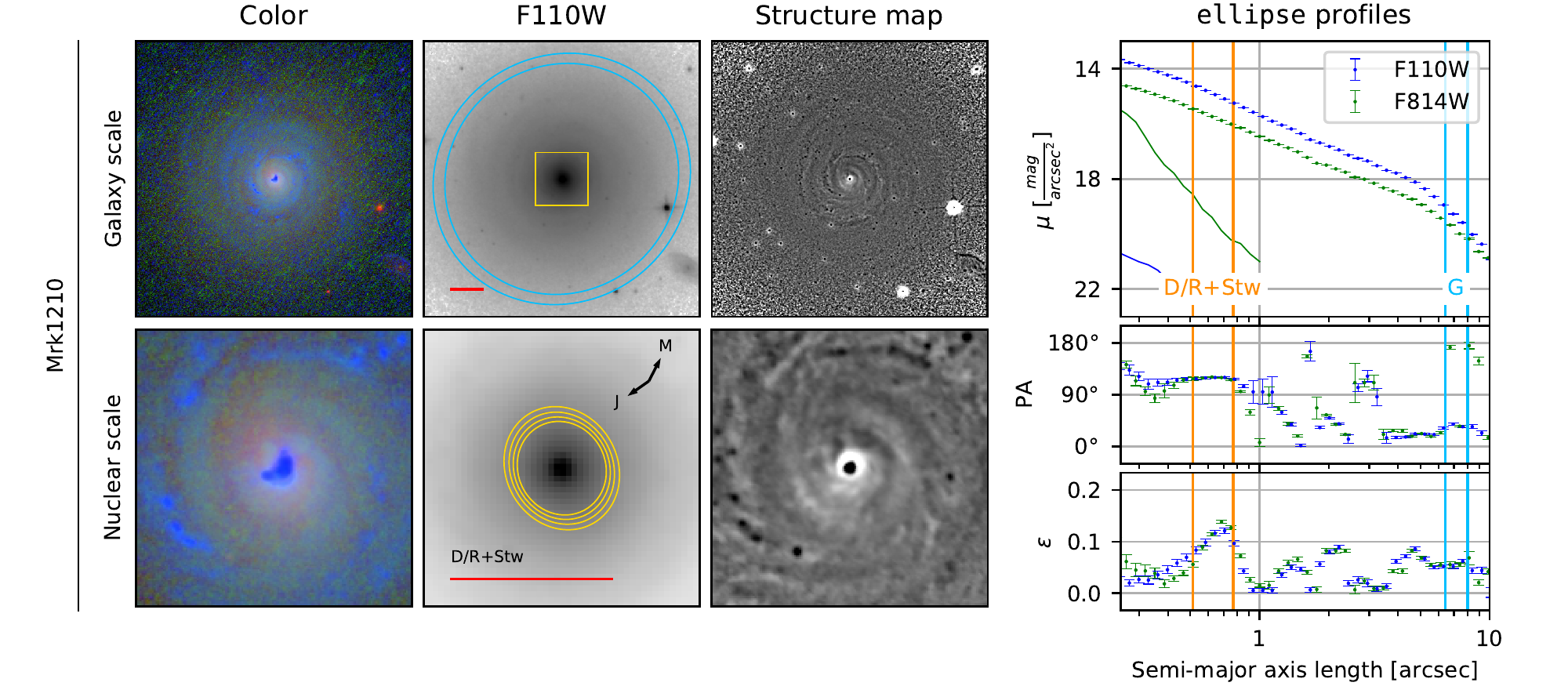}}
  \caption{Images and \texttt{ellipse} fits of Mrk~1210. The order of images and plots, as well as the meaning of symbols and designations are the same as in Fig.~\ref{fig:3rdPage1}.}
  \label{fig:3rdPage_Mrk1210}
 \end{figure}
 
 \begin{figure}[H]
  \centering
  \makebox[\textwidth]{\includegraphics[scale=1.]{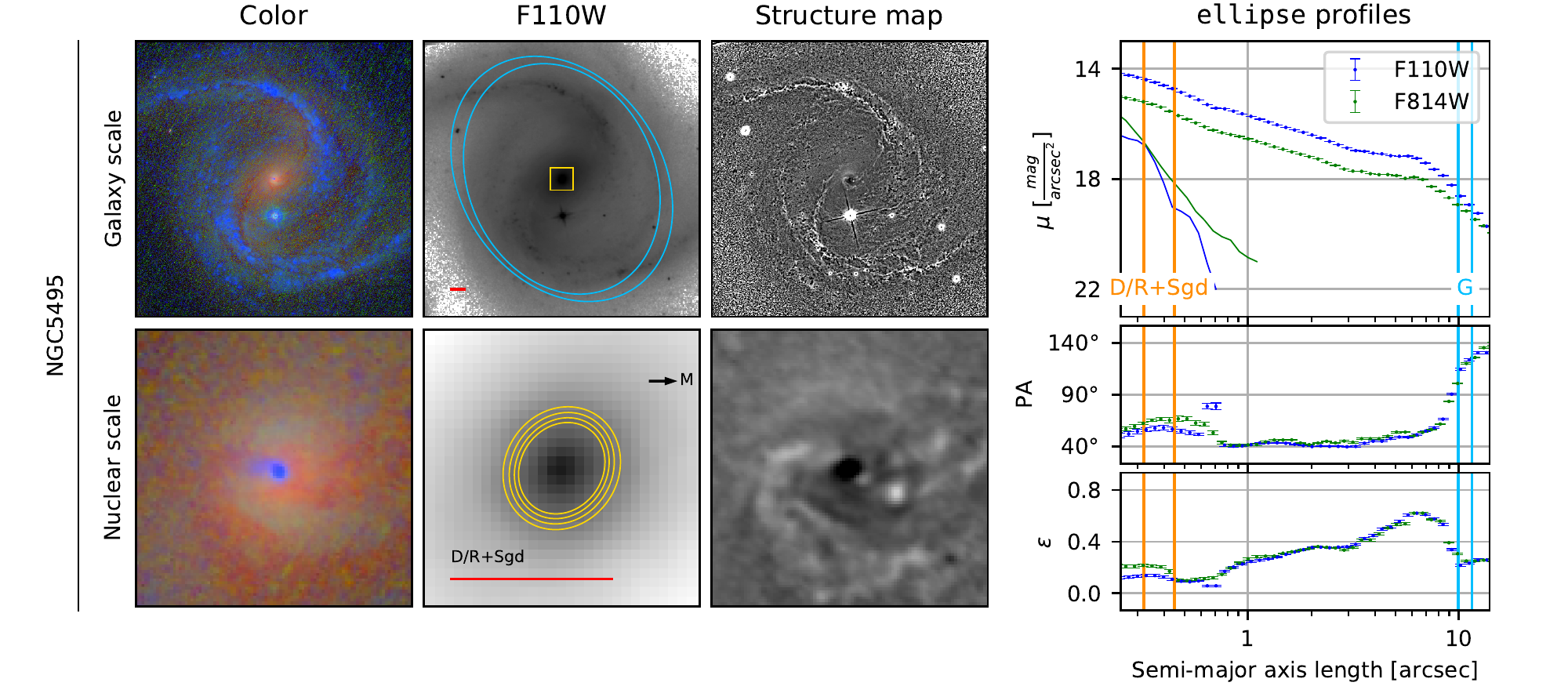}}
  \caption{Images and \texttt{ellipse} fits of NGC~5495. The order of images and plots, as well as the meaning of symbols and designations are the same as in Fig.~\ref{fig:3rdPage1}.}
  \label{fig:3rdPage_NGC5495}
 \end{figure}
 
 \subsection{Mrk~1210}
  
 We classify Mrk~1210 (the Phoenix Galaxy) as a face-on Sc galaxy. It has been discovered to be a Seyfert galaxy by \citet{1983Balzano} and its luminosity distance is $59.5$~Mpc \citep{2000Mould}. A jet position angle has been measured by \cite{2010Xanthopoulos} to be $125\degree$ at 50~pc. The spiral structure is wound clockwise with two arms clearly visible. We identify a $\sim 0.6$~arcsec ($170$~pc) nuclear feature characterized by a flattening in the position angle profile and a maximum in ellipticity. No dust features seem to be associated with it in the galaxy's structure map. However, spiral structure appears to be visible around it, so we find it likely to be either a disk or a ring. We note that the maser interpretation of the radio features seen in Mrk~1210 is less certain than in the other sources.
 
 \newpage
 
 \subsection{NGC~5495}
  
  NGC~5495 is a barred Sc galaxy (adapted from \citealt{1991Vaucouleurs}) with a luminosity distance of $97.5$~Mpc \citep{2000Mould}. It is moderately inclined and hosts a large bar with two bright, counterclockwise wound spiral arms. The nuclear region's radius is $\sim 0.4$~arcsec ($170$~pc), as indicated by flattened maxima in position angle and ellipticity profiles. A bar or spiral structure appears to be visible in dust within this nuclear region, so we classify it as either a disk or a ring. A two-arm grand-design nuclear spiral is also present (see discussion in Sect.~\ref{sect:morphology}).
 
 \begin{figure}
  \centering
  \makebox[\textwidth]{\includegraphics[scale=1.]{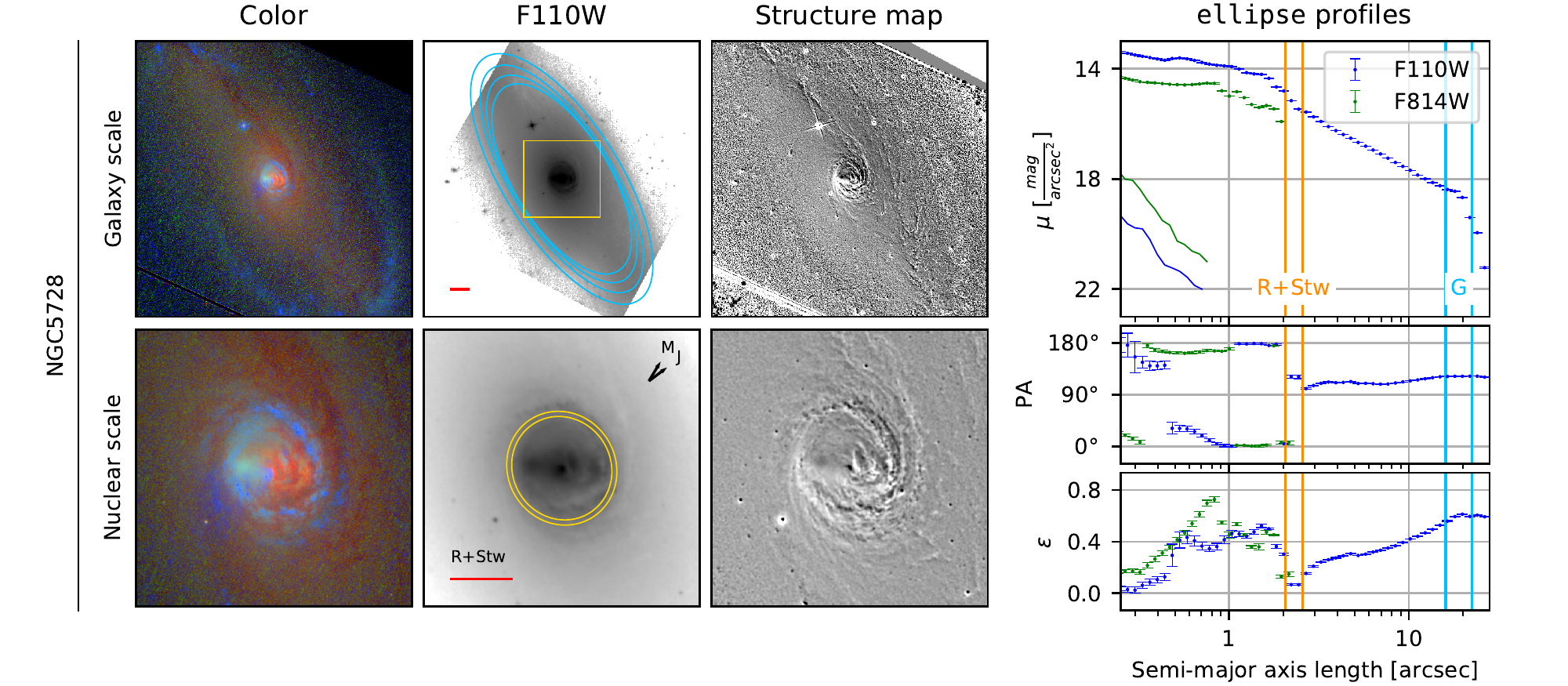}}
  \caption{Images and \texttt{ellipse} fits of NGC~5728. The order of images and plots, as well as the meaning of symbols and designations are the same as in Fig.~\ref{fig:3rdPage1}.}
  \label{fig:3rdPage_NGC5728}
 \end{figure}
  
  \subsection{NGC~5728}
  
  NGC~5728 is a barred Sa galaxy \citep{1991Vaucouleurs}, exhibiting Sy$1.9$ activity \citep{2006VeronCetty}. Its luminosity distance is $41.9$~Mpc \citep{2000Mould}. It is moderately inclined ($i\sim65\degree$) and contains a large bar, from which two counterclockwise-wound spiral arms extend. The position angle of a jet at 900~pc has been measured by \citet{1988Schommer} to be $307\degree$. We note that the maser interpretation of the radio features seen in NGC~5728 is less certain than in the other sources.
  
  In the case of NGC~5728 we identify a nuclear ring $\sim 2$~arcsec ($460$~pc) in radius (described in detail by \citealt{1988Schommer} and \citealt{1993Wilson}) as the nuclear region. The \texttt{ellipse} profile shows a discontinuity in position angle and near-zero ellipticity at this scale. This is not the smallest resolved structure in the galaxy images -- more features can clearly be seen inside it. However, it appears to be the smallest structure in the nucleus with a clear interpretation. The angular momentum of the nuclear ring is pointed toward $121\degree$ or $301\degree$. The nuclear ring is inclined at $\pm 21\degree$ with regard to the Celestial Sphere. Note that the \texttt{ellipse} data allow for four possible angular momentum orientations for the ring, see Sect.~\ref{sect:orient}. Outside the nuclear ring, a tightly-wound nuclear spiral can be seen. While it is undetected in the \texttt{ellipse} profiles, it is obvious in the structure map, demonstrating the advantage of using multiple tools to interpret the features in galactic nuclei. As noted above, more structure can be seen inside the nuclear ring of NGC~5728. While in the F110W image it closely resembles a bar along the East-West line connected with the nuclear ring, on the dust structure map it looks like a flocculent spiral with a smoother patch at the Eastern side in the blue filter, seen as a cone-like region in the color image. This feature has been identified by \cite{1988Schommer} and later confirmed by \cite{1993Wilson} to be an AGN ionization cone. The emission from the NLR overwhelms any dust structure in the same part of the image as the cone. This is why the inner nuclear spiral can only be seen on the western side of the structure map. Interestingly, the ionization cone corresponds to a region with positive radial velocity with respect to systemic, as shown using H$\alpha$ spectrometry by \citet[][see their Fig.~9]{1988Schommer}. If the ionization region corresponds to an outflow (as we would expect), it would have to be directed away from us. However, as argued by \cite{1993Wilson}, the true orientation of this main ionization cone (and its counterpart in the SW direction) is uncertain. \cite{1988Schommer} also report a jet in NGC~5728. It extends in the opposite direction from the large (SE) ionization cone and its projection is aligned with the projection of the megamaser disk's angular momentum.
  
  \newpage
 
 \begin{figure}
  \centering
  \makebox[\textwidth]{\includegraphics[scale=1.]{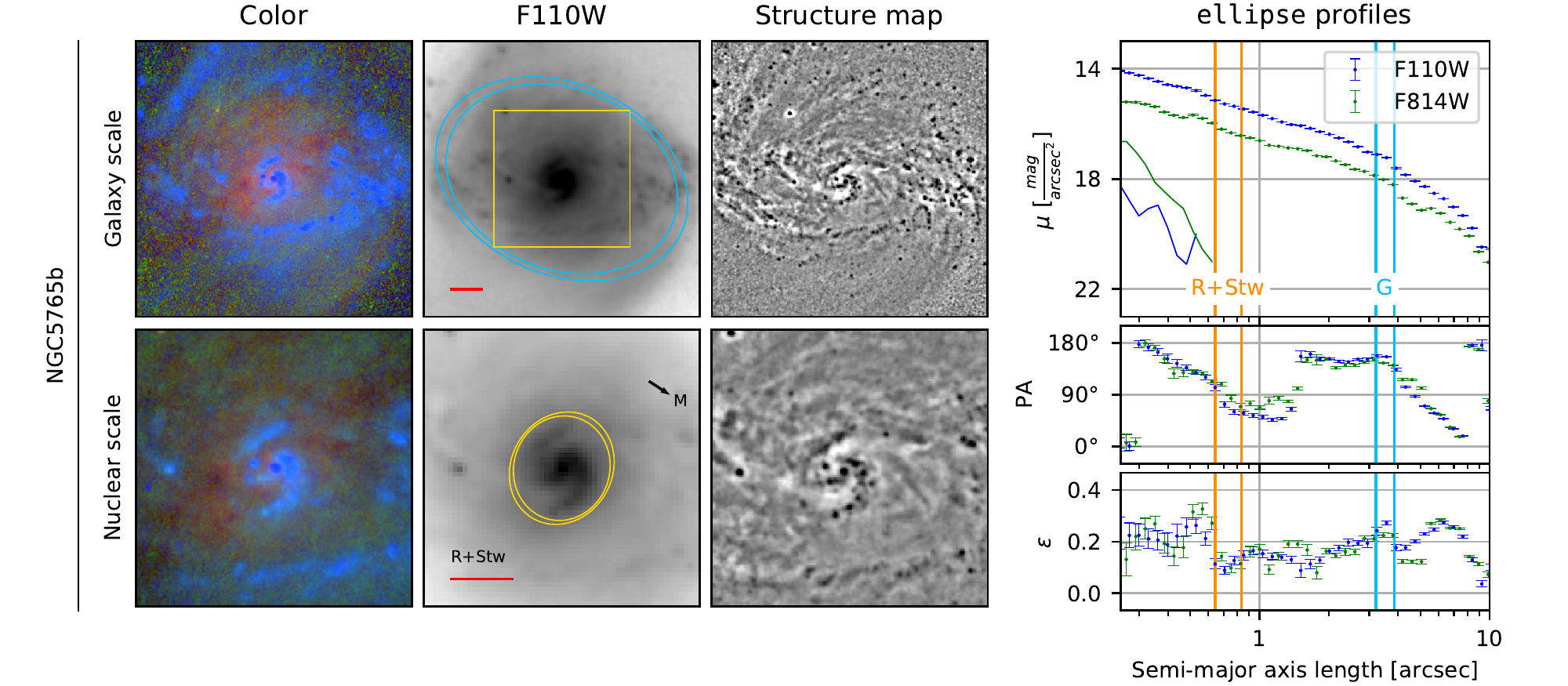}}
  \caption{Images and \texttt{ellipse} fits of NGC~5765b. The order of images and plots, as well as the meaning of symbols and designations are the same as in Fig.~\ref{fig:3rdPage1}.}
  \label{fig:3rdPage_NGC5765b}
 \end{figure}
  
  \subsection{NGC~5765b}
  
  NGC~5765b is an Sab galaxy located 126.3~Mpc from the Earth (\citealt{2016Gao}, angular diameter distance). It constitutes a pair with NGC~5765a and its activity has been classified as Sy2 \citep{2012Shirazi}. The galaxy contains two large-scale rings $\sim3.5$ and $\sim 1.5$~arcsec in radius (see Fig.~\ref{fig:3rdPage_NGC5765b}). Spiral structure is seen both within and outside each of the rings, it is wound clockwise with pitch angles changing between the rings (as can be seen in position angle \texttt{ellipse} profiles in Fig.~\ref{fig:3rdPage_NGC5765b}). Beyond the outer large-scale ring, the galaxy starts to exhibit signs of interaction with NGC~5765a. As the nuclear region we identify a small-scale ring with radius of $\sim 0.7$~arcsec ($\sim450$~pc). A nuclear spiral is clearly visible around it. While in the F110W image the spiral inside the nuclear ring appears grand-design, the structure map reveals rich structure associated with it. We therefore classify it as a tightly-wound spiral.
  
  \subsection{UGC~3193}
  
  The galaxy morphology for UGC~3193 is barred-Sab \citep{1991Vaucouleurs} and it is located $61.3$~Mpc from Earth \citep{2000Mould}, seen close to edge-on. It is a member of a galaxy group. The spiral structure is clear, with two large-scale arms wound clockwise. A large-scale bar is identified in the \texttt{ellipse} profiles at $\sim 3''$. The nuclear region we identify shows a clear maximum in ellipticity profile at $r\sim0.8$~arcsec ($220$~pc), corresponding to a minimum in position angle. A distinct feature is also visible in the surface brightness \texttt{ellipse} profile. While there appears to be a dust structure at the selected region, it does not straightforwardly point at its morphological type. However, as the \texttt{ellipse} fits contain prominent signatures of the flattened character of that structure, we treat it as a possible disk feature. On smaller scales, the \texttt{ellipse} profiles point at the existence of a bulge (minimum of ellipticity at $\sim 0.5$~arcsec) and a small-scale bar (maximum of ellipticity at $\sim 0.4$~arcsec), visible as a dark lane across the chosen nuclear region in the structure map.
  
  \subsection{UGC~6093}
  
  UGC~6093 is a barred-Sbc \citep{1991Vaucouleurs} with a luminosity distance of $158$~Mpc \citep{2000Mould}. The large-scale structure consists of a central bulge and a bar connected to a large-scale ring, from which two large spiral arms extend counterclockwise. As the central structure we identify a region with a minimum in the F814W ellipticity profile and a change in position angle opposite for F110W and F814W at $\sim 0.2$~arcsec ($150$~pc). No additional structure is visible in the brightness profile. While there appears to be a near-circular feature visible in the structure map, it is nearing on the PSF scale, so we classify this region as an uncertain bulge. Additionally, a dim two-arm grand-design spiral may be distinguished outside this region in the structure map. It is, however, too uncertain to be included in the region's classification.
  
  \newpage
 
 \begin{figure}
  \centering
  \makebox[\textwidth]{\includegraphics[scale=1.]{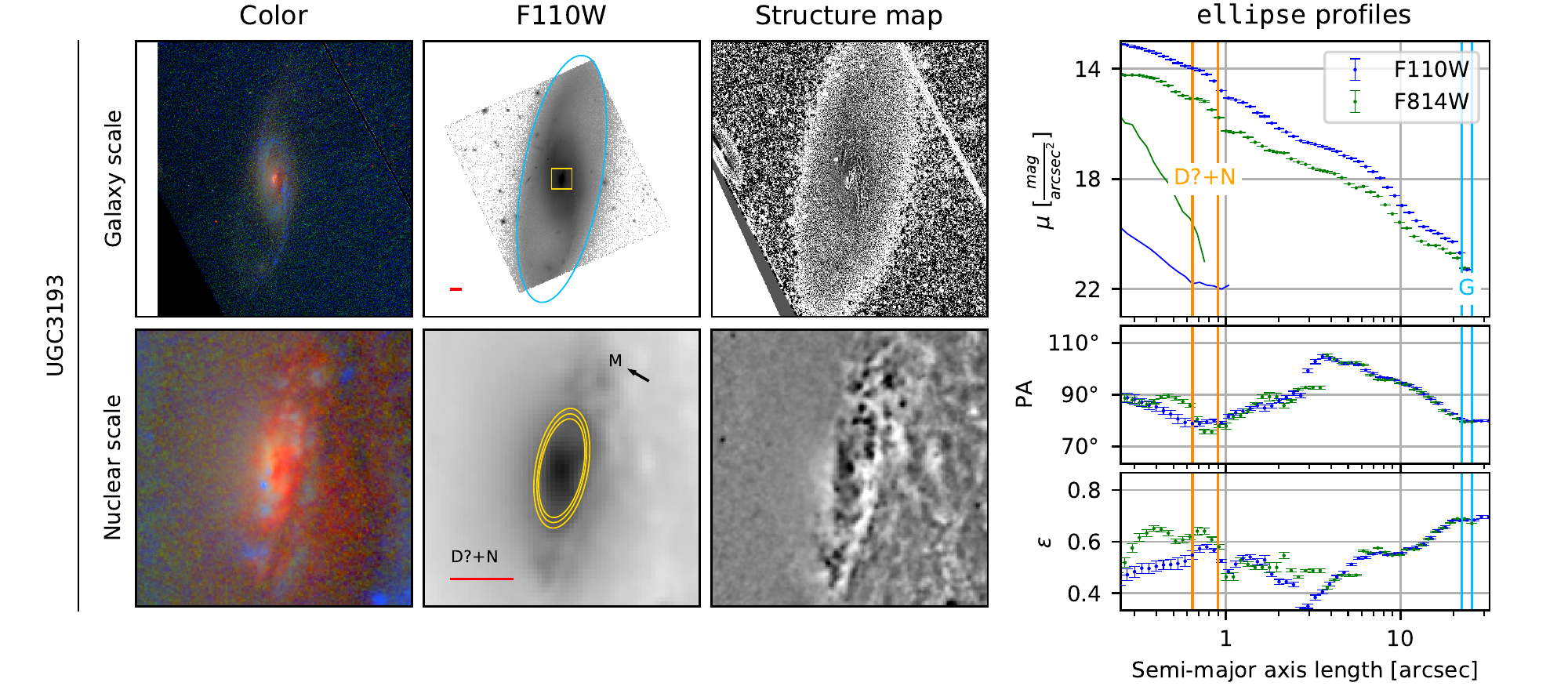}}
  \caption{Images and \texttt{ellipse} fits of UGC~3193. The order of images and plots is the same as in Fig.~\ref{fig:3rdPage1}. In the nuclear scale F110W image of UGC~3193 the red bar is $0.5$~kpc long in projected distance. The meaning of the remaining symbols and designations is the same as in Fig.~\ref{fig:3rdPage1}.}
  \label{fig:3rdPage_UGC3193}
 \end{figure}
 
 \begin{figure}
  \centering
  \makebox[\textwidth]{\includegraphics[scale=1.]{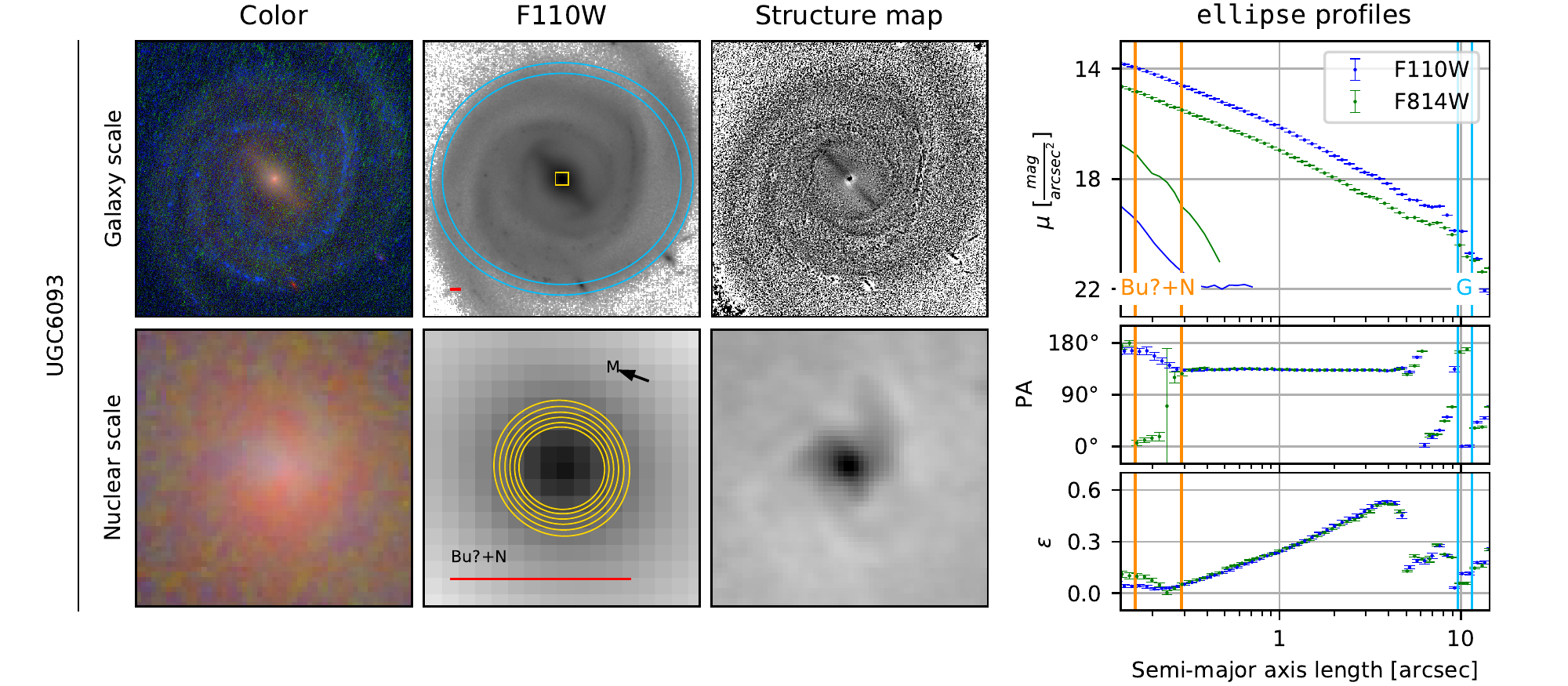}}
  \caption{Images and \texttt{ellipse} fits of UGC~6093. The order of images and plots, as well as the meaning of symbols and designations are the same as in Fig.~\ref{fig:3rdPage1}.}
  \label{fig:3rdPage_UGC6093}
 \end{figure}
 
 \section{Structure classification -- dependence on galaxy distance and scale of the nuclear region}\label{sect:classVsDist}
 
 \setcounter{figure}{0}
 \renewcommand{\thefigure}{B\arabic{figure}}
 
 \subsection{Dependence on galaxy distance}
 
\begin{figure}
 \centering
 \begin{tabular}{cc}
  \includegraphics{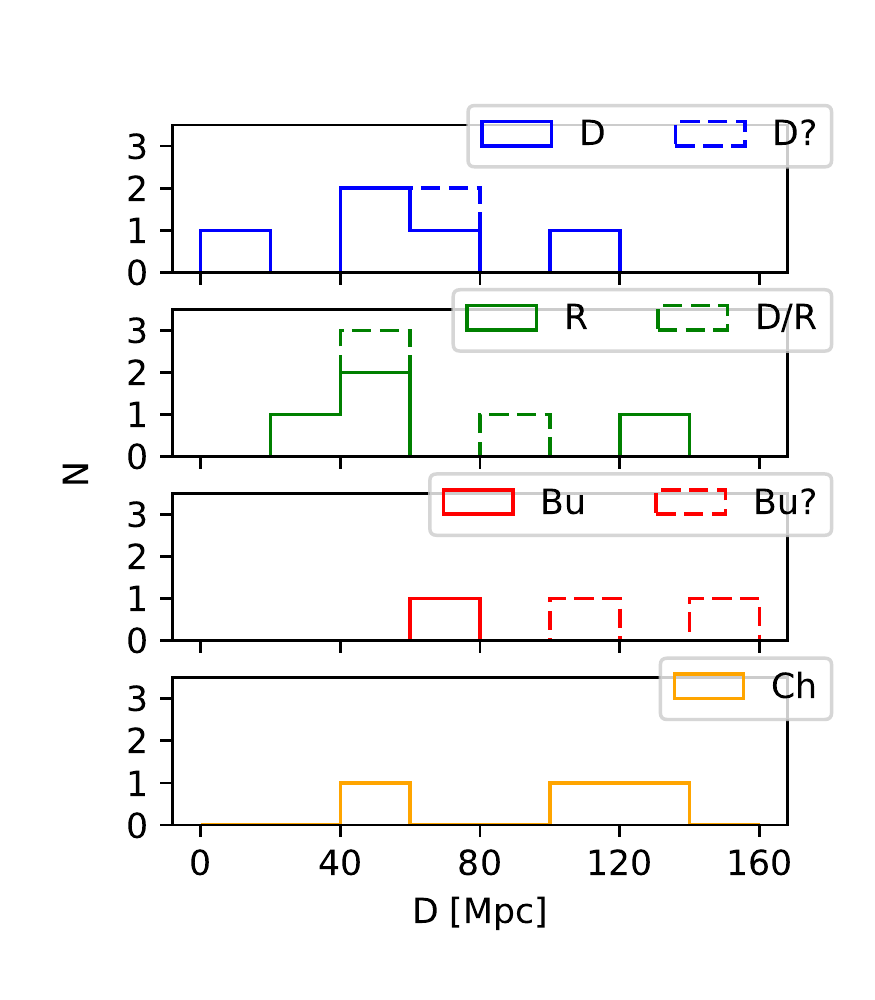} & \includegraphics{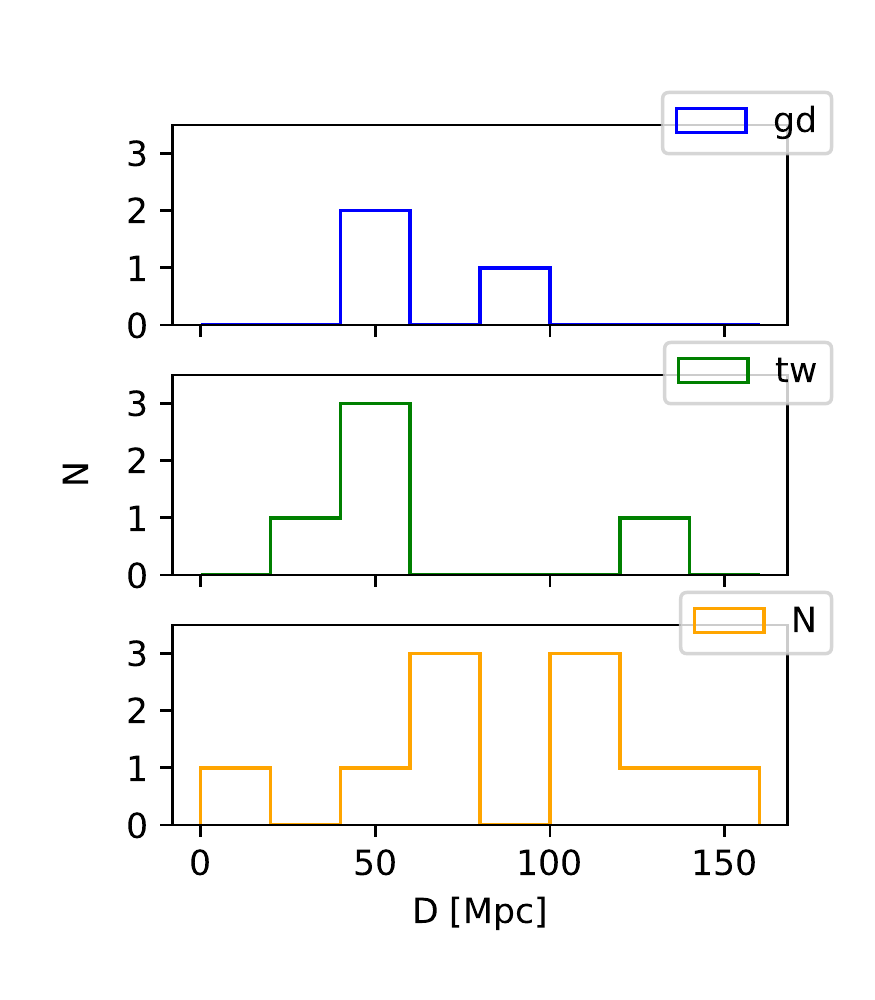}
 \end{tabular}
 \caption{Left: distribution of nuclear region classes as a function of angular diameter distance to the host galaxy. Notation: D~--~disk, R~--~ring, Bu~--~bulge, Ch~--~chaotic nuclear dust structure, ``?'' denotes uncertain classification. Right: distribution of nuclear spiral classes as a function of angular diameter distance to the host galaxy. Possible nuclear spiral classes: gd~--~grand-design nuclear spiral, tw~--~tightly-wound nuclear spiral, N~--~no nuclear spiral visible.}
 \label{fig:distDists}
 \end{figure}

 We assess the robustness of our classification by analyzing how the classes depend on distance to the host galaxy (and, correspondingly, the image resolution). The distribution of each class as a function of host galaxy distance is shown in Fig.~\ref{fig:distDists}. The histograms on the left depict the nuclear region by class. Disky and ring-like nuclear structures are best visible for close galaxies due to superior resolution, allowing us to resolve features associated with the flatness of a structure, such as nuclear spirals. Further away, more nuclear regions are classified as ``chaotic'' -- some of which are possibly disky structures that are too distant to allow certain classification -- or ``possible bulges''. Note that due to our simplistic approach to their identification, bulge classification is treated by us with caution. To properly decompose bulges from nuclear disks requires 2D image fitting (e.g., \citealt{2016Laesker}), which is beyond the scope of this work. The main reason to include nuclear bulges in our analysis is to avoid measurement of orientation of non-disky nuclear regions (see Sect.~\ref{sect:orient}) -- a goal that should be achieved even if some of the Bu?\ objects are misclassified. The dependencies above show that there are biases in our classification due to the decrease in available image resolution with distance, as signaled by the ``unsure'' (question marks) or ``chaotic'' (Ch) categories. Similar reasoning holds true for the nuclear spirals, which we appear to assign as ``N'' if unresolved (see the right panel of Fig.~\ref{fig:distDists}).
 
 In the left part of Fig.~\ref{fig:distDists} one can note a division between $D\lesssim 80$~Mpc, where almost all the classifications are robust, and $D\gtrsim 80$~Mpc, with a considerable fraction of either unsure (marked with ``?'') or chaotic (Ch) nuclear regions. We therefore conclude that our classifications for galaxies less than $80$~Mpc away are not affected by angular resolution bias and can be specifically trusted. Additionally, we see that while for galaxies beyond $D\sim60$~Mpc our chosen nuclear regions assume different morphologies, almost all of them are disk-like (either of D or R type) below that limiting distance. 
 
 In the right-side histograms of Fig.~\ref{fig:distDists} the distribution in host distance for the classes of nuclear spirals is shown. Grand-design (gd) structures are easily recognizable in the entire range of distances and it appears unlikely that any of the ``N'' (no nuclear spiral visible) sources are in fact unresolved ``gd''s (grand-design nuclear spirals). While the ``N'' sources in further galaxies must in part be unresolved ``tw''s, their flattened distribution with distance suggests that there are intrinsically chaotic dust structures in our sample as well.
 
 % \textcolor{red}{We would then expect the ratio of the number of both tightly wound (tw) and None (N) nuclear spirals to the number of 'gd' nuclear spirals ([tw+N]/gd) to be constant with distance (as some 'tw' would be seen as 'N' further away). However, this is not the case -- the relative number of grand-design spirals seems to increase with distance. One might suspect this to be a selection effect, which would connect grand-design nuclear spirals with increased luminosity of megamaser emission and / or nuclear activity. While such a connection does not seem to be supported by previous studies (\citealt{2003Martini}, but see also \citealt{2014Davies}), \cite{2016GreeneSMBHmass} suggests that megamaser galaxies harbour SMBH of masses lower than in the general population. It is possible that a lower-mass SMBH could facilitate the formation of nuclear grand-design spiral. However, the small number of galaxies in our sample urges skepticism concerning this hypothesis. A change in classification of only one of the grand-design spirals in galaxies further than $100$~Mpc to either 'N' or 'tw' would make the ratio [tw+N]/gd roughly constant with distance.}
 
 \begin{figure}
 \centering
 \begin{tabular}{cc}
  \includegraphics{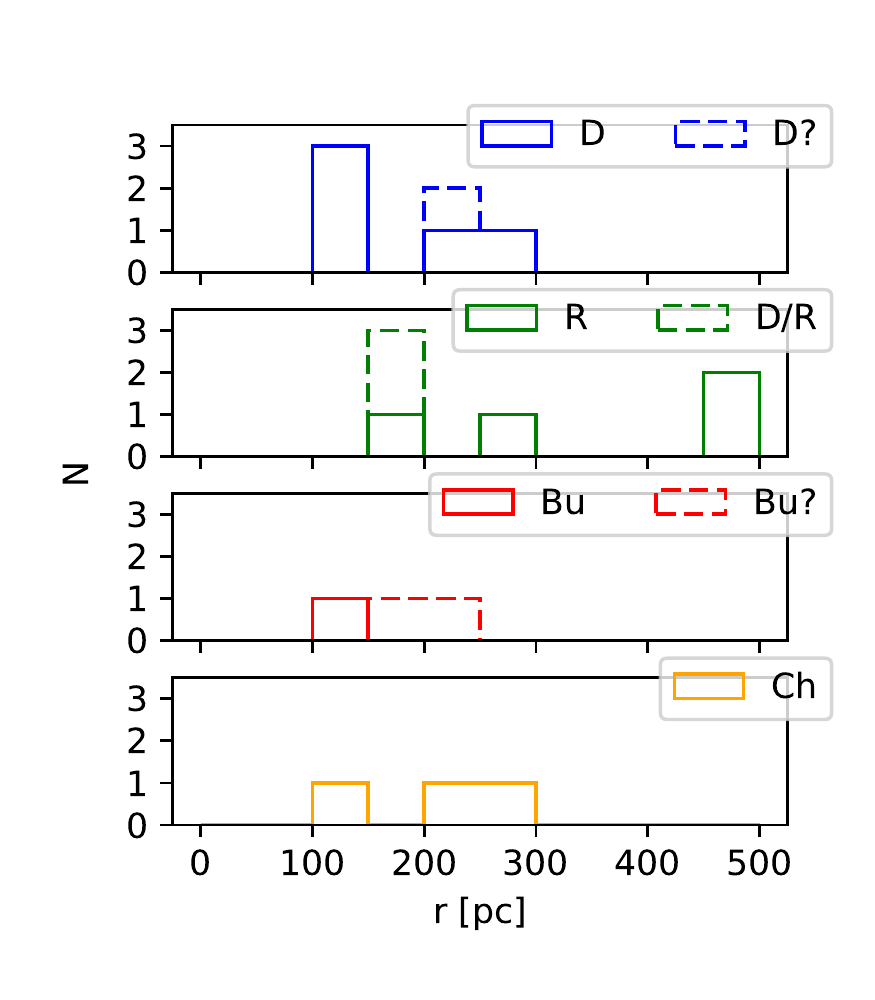} & \includegraphics{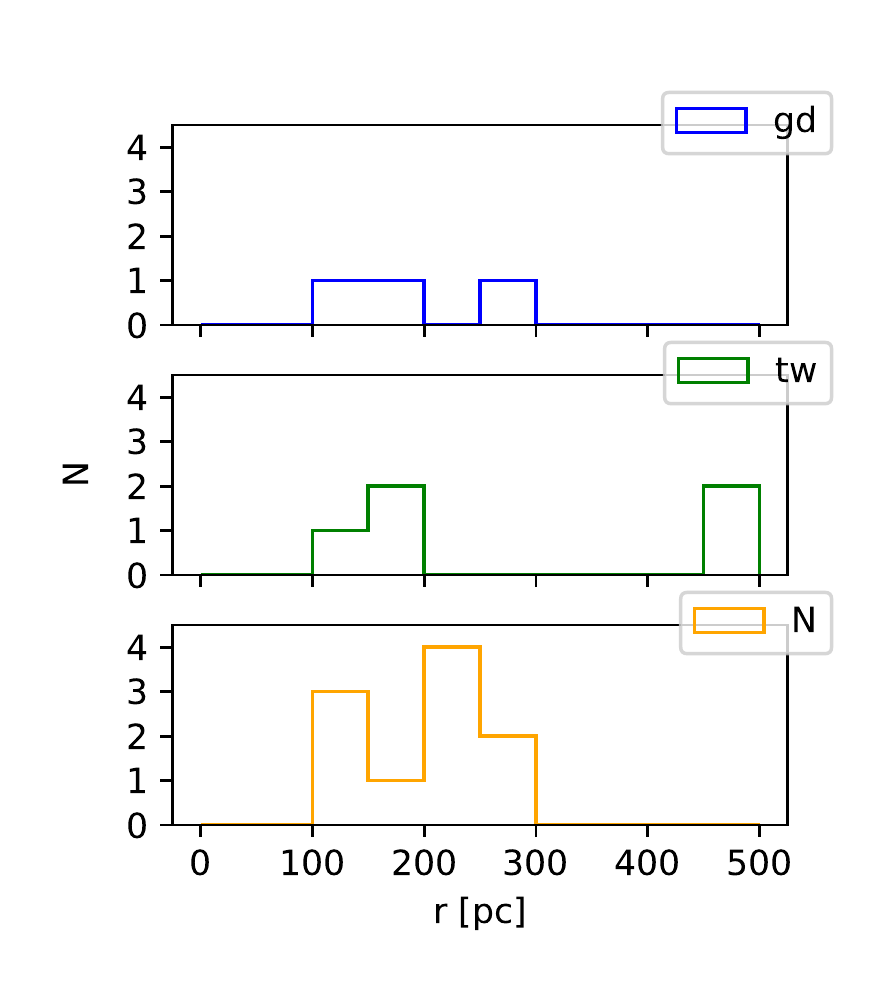}
 \end{tabular}
 \caption{Left: distribution of nuclear region classes as a function of the region scale. Notation: D~--~disk, R~--~ring, Bu~--~bulge, Ch~--~chaotic nuclear dust structure, ``?'' denotes uncertain classification. Right: distribution of nuclear spiral classes as a function of the scale of the corresponding nuclear region. Nuclear spiral classes: gd~--~grand-design, tw~--~tightly-wound, N~--~no nuclear spiral visible.}
 \label{fig:distScales}
 \end{figure}
 
 \subsection{Dependence on scale of the nuclear region}
 \label{sect:morph_scale}
 
 %We first discuss the morphology of the nuclear regions (Table~\ref{tab:orient}). The radial scales of these regions (column 13 of Table~\ref{tab:orient}) range from $r\sim 30$~pc (NGC2960) to $r\sim 460$~pc (NGC5728) depending on the size of the smallest resolved nuclear structure, with most values clustered around $150$~pc (see Fig.~\ref{fig:distScales}). 
 
 %\textcolor{red}{We have classified the type of the region within size $\sim r$ (Sect.~\ref{sect:identNuclStruct}) as a disk (D), ring (R), bulge (Bu), bar (B) or having no discernible character (Ch). We have also looked for nuclear spirals, which we further characterize as being tightly-wound (tw) or grand-design (gd). For more details on the classification scheme see Sect.~\ref{sect:classification}.}
 
 The left-side histograms of Fig.~\ref{fig:distScales} show the distribution of different classes of nuclear regions as a function of the physical scale of the region. We observe a transition between a mixture of nuclear structure types at $150$--$300$~pc to almost exclusively disky environments at smaller scales. We observe disk-like nuclear regions in virtually \textit{all} of our galaxies closer than $\sim 60$~Mpc (see Fig.~\ref{fig:distDists}). Perhaps in some of the more distant sources similar features remain undetected at larger scales available to us. Distributions of nuclear spiral types with scale seem relatively flat, suggesting that the type of such structure does not depend on the physical scale of the region.

 As noted in Sect.~\ref{sect:identNuclStruct}, all of our nuclear structures were identified at sizes of at least $0.3$~arcsec. To make sure that such limiting angular size is sufficient and resolution effects do not affect the results presented in Sections~\ref{sect:morphology} and~\ref{sect:results_angmom}, we have re-derived these results with a more restrictive limit on angular size of a nuclear region of $0.5$~arcsec. As a result, we limited our sample to 10 of the 18 galaxies. We used the same limit between small and large nuclear regions of $200$~pc to divide the re-derived sample. Due to different distances to galaxies, the sizes of these sub-samples were still comparable, with 5~small and 5~large nuclear regions, respectively. As can be seen in results of KS tests presented in Table~\ref{tab:KSlimited}, our conclusions (see Sections~\ref{sect:morphology} and~\ref{sect:results_angmom}) remained mostly unchanged (cf. Table~\ref{tab:KS}). We note two differences in comparison with Table~\ref{tab:KS}. Firstly, in this limited sample we cannot rule out that both small and large nuclear regions align with the kpc-scale galaxy, although the p-value for random orientation of small nuclear regions is still larger than that for their being aligned (see discussion in Sect.~\ref{subsec:galreg}). Secondly, in this test the distribution of relative position angles between large nuclear regions and the masing disks is seemingly inconsistent with random (p-value $=0.01$), while the small nuclear regions remain randomly oriented with regard to their megamasers. We find this unlikely to be a physical effect, as the information about maser orientation would have to ``jump over'' the small nuclear regions. Moreover, the results of KS tests using the 3D~angles between angular momenta of large nuclear regions and masing disks remain consistent with their random relative orientation (with p-value even improving from $0.42$ in the initial sample, see Table~\ref{tab:KS}, to $0.64$ in the limited one, Table~\ref{tab:KSlimited}). Since these results utilize additional information, the inclination of nuclear regions, we regard them as more trustworthy than the position angles alone. We conclude that even in the sample limited to $0.5$~arcsec both small and large nuclear regions are consistent with being fully randomly oriented with respect to their masing disks.
 
 %It is possible that both distributions of $\Delta$PA follow a more complex distribution than tested here and our result is an artifact resulting from trying to fit too simplified a distribution.
 
 \begin{table}
  \centering
  \begin{threeparttable}
  \caption{Results of the KS tests\\for nuclear regions larger than $0.5$~arcsec.}
  \renewcommand*{\arraystretch}{1.1}
  \begin{tabular}{|cc|c|c|c|cc|}
    \hline
    \multicolumn{2}{|c|}{Species} & Angle & Limit & N & KS-stat & p-value \\
    (1) & (2) & (3) & (4) & (5) & (6) & (7) \\
    \hline
    Gal & Nuc & $\Delta$PA & $30\degree$ & 10 & $0.30$ & $0.27$ \\ 
Gal & Nuc & $\Delta$PA & $90\degree$ & 10 & $0.42$ & $0.04$ \\ 
Gal & Nuc(S) & $\Delta$PA & $30\degree$ & 5 & $0.50$ & $0.11$ \\ 
Gal & Nuc(S) & $\Delta$PA & $90\degree$ & 5 & $0.32$ & $0.59$ \\ 
Gal & Nuc(L) & $\Delta$PA & $30\degree$ & 5 & $0.43$ & $0.23$ \\ 
Gal & Nuc(L) & $\Delta$PA & $90\degree$ & 5 & $0.64$ & $0.02$ \\ 
\hline
Nuc & Mas & $\Delta$PA & $90\degree$ & 10 & $0.42$ & $0.04$ \\ 
Nuc(S) & Mas & $\Delta$PA & $90\degree$ & 5 & $0.32$ & $0.59$ \\ 
Nuc(L) & Mas & $\Delta$PA & $90\degree$ & 5 & $0.69$ & $0.01$ \\ 
Nuc & Mas & 3D & $90\degree$ & 10 & $0.31$ & $0.24$ \\ 
Nuc(S) & Mas & 3D & $90\degree$ & 5 & $0.32$ & $0.61$ \\ 
Nuc(L) & Mas & 3D & $90\degree$ & 5 & $0.31$ & $0.64$ \\ 
\hline
Gal$^*$ & Mas & $\Delta$PA & $90\degree$ & 10 & $0.30$ & $0.27$ \\ 
Gal$^*$ & Mas & 3D & $180\degree$ & 10 & $0.30$ & $0.26$ \\ 
\hline

  \end{tabular}
\begin{tablenotes}\footnotesize
 \item \textbf{Notes:} Results of one-sample KS tests for a sample limited to galaxies with nuclear regions larger than $0.5$~arcsec. The division between small and large nuclear regions remains at $200$~pc. For description of KS tests see Sect.~\ref{sect:dissRelOrient}. The meaning of columns and notation is the same as in Table~\ref{tab:KS}.
 
 %Columns~1-2: structures in the galaxy whose angular momenta orientations are being compared. Gal -- angular momentum of the galaxy as a whole, Gal$^*$ -- the same limited to galaxies with orientation fixed using rotation curves or the relative dust lane prominence method, Nuc -- the nuclear regions (in some cases divided into S -- those with $r<200$~pc and L -- with $r>200$~pc), Mas -- the megamaser disk. Column~3: type of relative angle: $\Delta$PA is the position angle difference (as defined in the text), 3D denotes the 3D angles between angular momenta. Column~4: upper limit of the range of the respective angle in which the control distribution is random (the lower limit in all cases is $0$). Column 5 -- number of sources in the sample. Columns~6-7 -- KS-statistic (column~6; maximal difference between CDFs, or Cumulative Distribution Functions, of the sample and test distribution) and p-values (column~7; likelihoods of the sample being drawn from the test distribution) resulting from each of the tests.
\end{tablenotes}
\label{tab:KSlimited}
\end{threeparttable}
\end{table}
 
\end{document}